\documentclass[traditabstract]{aa} 
\usepackage{graphicx}
\usepackage{txfonts}

\usepackage{ulem}

\def\ms{\hbox{\,m\,s$^{-1}$}}         
\def\cms{\hbox{\,cm\,s$^{-1}$}}       
\def\m2s2{\hbox{\,m$^{2}$\,s$^{-2}$}} 
\def\kms{\hbox{\,km\,s$^{-1}$}}       

\def\Mjup{\hbox{$\mathrm{M}_{\rm Jup}$}}

\def \1s{$1\,\sigma$}

\def \t0{T$_0$}

\def\modif{}
\def\modiff{}

\begin{document}


\author{
F. Bouchy  \inst{1,2}
\and R.F. D\'iaz \inst{1,2}
\and G. H\'ebrard \inst{1,2}
\and L. Arnold \inst{2}
\and I. Boisse \inst{3}
\and X. Delfosse \inst{4}
\and S. Perruchot \inst{2}
\and A. Santerne \inst{5}
}

\institute{
Institut d'Astrophysique de Paris, UMR7095 CNRS, Universit\'e Pierre \& Marie Curie, 
98bis Bd Arago, 75014 Paris, France\\
\email{bouchy@iap.fr}
\and
Observatoire de Haute-Provence, CNRS/OAMP, 04870 St Michel l'Observatoire, France
\and
Centro de Astrof\'isica, Universidade do Porto, Rua das Estrelas, 4150-762 Porto, Portugal
\and
UJF-Grenoble 1 / CNRS-INSU, Institut de Plan\'etologie et d'Astrophysique de Grenoble, 
UMR 5274, Grenoble, 38041, France
\and 
Aix-Marseille Universit\'e, CNRS, Laboratoire d'Astrophysique de Marseille,  UMR 7326, 13388 Marseille, France
}

   \title{$SOPHIE+$: First results of an octagonal-section fiber \\ for high-precision radial 
   velocity measurements}


   \date{Received ; accepted }

  \abstract
  {High-precision spectrographs play a key role in exoplanet searches and Doppler asteroseismology 
  using the radial velocity technique. The 1 {\ms} level of precision requires very high stability and uniformity of 
  the illumination of the spectrograph. In fiber-fed spectrographs such as SOPHIE, the fiber-link scrambling properties 
  are one of the main conditions for high precision. To significantly improve the radial velocity precision of the SOPHIE 
  spectrograph{\modif , which was limited to 5-6 {\ms},} we implemented a piece of octagonal-section fiber in the fiber link. We present here the scientific 
  validation of the upgrade of this instrument, demonstrating a real improvement. The upgraded instrument, renamed 
  SOPHIE+, reaches radial velocity precision in the range of 1-2 {\ms}. It is now fully efficient for the detection of 
  low-mass exoplanets down to 5-10 M$_{\oplus}$ and for the identification of acoustic modes down to a few tens 
  of {\cms}.}

  \titlerunning{SOPHIE+: First results}
  \authorrunning{F. Bouchy et al.}

   \maketitle


\section{Introduction}

In the last few decades, the continuous improvement of radial velocity (RV) measurements has revolutionized 
exoplanetology. Doppler measurements have illustrated their capabilities by extending the exoplanet search 
around a wide variety of stars, opening the possibility of exploring the domain of low-mass planets down to 
a few Earth masses. This technique led to the discovery and characterization of multiple planetary systems, perform 
long-term surveys to find Jupiter analogs, establish the planetary nature, and characterize the mass, 
eccentricity, and obliquity of transiting planets. Using the same instruments, Doppler asteroseismology on bright 
stars has also permitted revelation of their tiny acoustic oscillations modes and probing of their internal structures.

One RV method makes use of a molecular absorption cell filled with iodine gas to impress lines 
of stable wavelength on the incoming starlight (e.g., Cochran \& Hatzes \cite{cochran94}; 
Walker et al. \cite{walker95}; Butler et al. \cite{butler96}). The stellar and reference spectra are then 
recorded from the same beam, thereby circumventing the problem of any stellar photocenter shift or instrumental 
instability. However, this approach considerably increases the photon-noise RV errors due to the limited spectral 
range (5000 - 6200 {\AA}) and the absorption of the iodine gas. Another RV method, which 
avoids this drawback, makes use of a fiber-optic feed for the starlight, plus a second fiber carrying light 
from a stable wavelength source (Brown et al. \cite{brown94}; Mayor \& Queloz \cite{mayor95}). This 
approach, {\modif which requires high opto-mechanical stability of the spectrograph, also}  strongly depends on the capability of the optical fiber to scramble and homogenize the stellar beam.

The SOPHIE spectrograph (Bouchy et al. \cite{bouchy06}; Perruchot et al. \cite{perruchot08}) has been in 
operation since October 2006 at the 1.93-m telescope of Observatoire de Haute-Provence (OHP). Benefiting 
from experience acquired on ELODIE (Baranne et al. \cite{baranne96}) and HARPS (Pepe et al. \cite{pepe02}), 
SOPHIE was designed to obtain high-precision RV measurements with much higher throughput than 
its predecessor. SOPHIE is a fiber-fed cross-dispersed, environmentally stabilized echelle spectrograph covering 
the visible wavelength domain from 3872 to 6943 {\AA}. It plays an efficient role in the search for northern 
extrasolar planets (e.g., Bouchy et al. \cite{bouchy09}; H\'ebrard et al. \cite{hebrard10}), and in the Doppler 
follow-up of photometric surveys for planetary transit searches, such as SuperWASP (Collier-Cameron et al. \cite{cameron07}), HAT (Bakos et al. \cite{bakos07}), CoRoT (Barge et al. \cite{barge08}) and {\it Kepler} (Santerne 
et al. \cite{santerne11}). SOPHIE is also used for planetary system obliquity measurements and has allowed the detection of the first two cases of spin-orbit misalignment (H\'ebrard et al. \cite{hebrard08}; Moutou et al. \cite{moutou09}). SOPHIE also plays a significant role in the detection identification of acoustic oscillation modes 
(p-modes) on solar-like stars (Mosser et al. \cite{mosser08}). However, the first years of operation showed that the RV precision obtained on stable stars was limited to 5-6 {\ms} in the best cases (e.g., Boisse et al. \cite{boisse09}; Diaz et al. \cite{diaz12}). Although well adapted for the detection and characterization of giant planets, 
this precision is far from being appropriate for the super-Earth and Neptune-like planets, which require the 1-2 {\ms} precision level. This RV limitation was identified as being mainly caused by the insufficient scrambling of the fiber link and the high sensitivity of the spectrograph to illumination variations. To significantly improve the Doppler precision, we implemented a new fiber link that included a piece of octagonal-section fibers and tested this concept on the sky for the first time. 

Section 2 is devoted to a description of scrambling properties of fibers commonly used on astronomical spectrographs. We report in section 3 the RV systematics measured on the SOPHIE spectrograph due to the insufficient scrambling of fibers. We then discuss in section 4 the properties of square- and octagonal-section fibers, which allow the scrambling gain of optical fibers to be considerably improved. In section 5 we present results of the scientific validation of the upgraded spectrograph, hereafter renamed SOPHIE+.

\section{Scrambling properties of fibers used for echelle spectrograph} 

A review of fibers in astronomy has been given by Heacox \& Connes (\cite{heacox92}). Fiber-fed spectrographs 
use multi-mode step-index fibers with core sizes typically in the 50- to 500-$\mu$m range. One significant 
characteristic of fibers is their ability to scramble the input image. The geometry of cylindrical 
fibers introduces two dimensions of scrambling, azimuthal, and radial. Both theory (Heacox \cite{heacox87}) 
and experiments (Hunter \& Ramsey \cite{hunter92}, Avila et al. \cite{avila06}, \cite{avila08}) show that fibers 
provide a high degree of azimuthal scrambling but an incomplete radial scrambling. 


The near-field pattern of a fiber is defined as the brightness distribution across its output face. In most cases, a spectrograph images this output face, which corresponds to the slit entrance, directly onto the detector as a function of wavelength. Variations in illumination on the entrance fiber will reflect themselves in a variation of the spot profile, which leads to RV shifts. The fiber-scrambling gain is defined as the ratio between the photocenter displacement of the near-field at the fiber ouput and the displacement of the input beam. 
Laboratory tests (e.g., Casse \cite{casse95}; Bouchy \& Connes \cite{bouchy99}; Avila et al. \cite{avila06}, \cite{avila08}) show that the motion of the output photocenter may be of the order of 100-200 times smaller than the input photocenter. 
The 1 {\ms} precision requires 
a stabilization of the star image on the fiber input {\modif at the level of few 0.01 arcsec}.
However, image stabilization (guiding and centering at the entrance fiber) is not the only factor that may introduce illumination variation. The image size, mainly driven by the seeing condition, telescope focusing, and image elongation due to atmospheric dispersion also plays a role in fiber illumination.  

The near-field pattern of the fiber-output face is not the only aspect to consider. One should also 
take carefully into account the far-field pattern, which is defined as the cross-section of the beam far from the output face 
or the angular distribution of light exiting the fiber. 
Far-field variations are projected onto the spectrograph pupil and cause changes in grating illumination. 
Spectrograph aberrations and grating imperfections may then shift the center of the light distribution on the 
detector as a function of the input light distribution and induce RV shifts. The amount of this effect strongly 
depends on the instrument aberrations, and it is not possible to give numbers that are 
valid in general. However, this aspect should be taken into account in the spectrograph optical design,  
which has to be as insensitive as possible to the pupil illumination variations 
or at least to be symmetrical along all spectral orders in order to be compensated in the RV measurements. 

On some fiber-fed spectrographs, the pupil of the telescope is imaged on the input face of the fiber 
(pupil injection). In that case, the near-field pattern is expected to be homogeneous (apart from the 
central obstruction of the telescope) and very stable. On the other hand, the far field sees the moving image  
of the telescope focal plan suffering from centering and guiding effects. This far-field pattern is only 
scrambled in azimuth by the fiber, but not radially. These changes in illumination are projected 
to the spectrograph pupil and grating. This is the case for spectrographs like 
FEROS (2.2-m ESO), HARPS in EGGS mode (3.6-m ESO), and FLAMES (VLT-ESO). On these instruments, 
the typical RV precision is not better than 20 {\ms} (e.g. Setiawan et al. \cite{setiawan04}; Loeillet et al. \cite{loeillet08}), 
and we strongly suspect that it comes from unstable illumination of the spectrograph grating due 
to the pupil injection mode in the fiber input. It is then more adapted to having image injection with a guiding system 
keeping the star well centered on the fiber entrance. In that case, the far field of the fiber sees 
the stable pupil of the telescope, which is projected on the spectrograph grating. 

To increase scrambling properties, one may incorporate a double-fiber scrambler (Brown et al. \cite{brown91}), 
in which a pair of fibers is coupled together using a pair of microlenses, separated by their common focal lengths. 
The fibers then see each other at infinity, causing the near- and far- fields to be interchanged. 
The near-field pattern is then projected onto the spectrograph pupil and the far-field pattern 
on the output face of the fiber, hence at the spectrograph slit entrance. The guiding errors on the fiber 
entrance are then almost not visible on near-field fiber output due to this field exchange. 
The scrambling gain of such a system was estimated both in the laboratory and on the sky 
to be between two and ten (Hunter \& Ramsey \cite{hunter92}, Brown et al. \cite{brown94}, Casse \cite{casse95}). 
This is the case for spectrographs like ELODIE (1.93-m OHP), SOPHIE in HR mode (1.93-m OHP), 
and HARPS in standard HAM mode (3.6-m ESO). The typical throughput of these double scramblers is 
about 85\%. 

Another solution proposed by Avila et al. (\cite{avila06}, \cite{avila08}) consists of inserting a mechanical scramblers, which squeeze the fiber with mini-bendings. Although it improves the scrambling gain by a factor up to eight, this solution significantly increases the focal ratio degradation and then reduces the throughput of the fiber link. 

As proposed by Connes et al. (\cite{connes96}), the use of a single-mode fiber acts as a perfect single-mode spatial filter. All cross-sections of the output beam are quasi-Gaussian and preserve no memory of the input beam geometry. Hence a single-mode fiber behaves as an ideal scrambler. It may be matched to the Airy pattern at the focus of a diffraction-limited telescope. On the ground, this solution is limited to a telescope pupil smaller than the Fried coherence length (Fried \cite{fried66}) which is roughly 12 cm in the visible for 1 arcsec seeing or requires adaptive optics. 

Connes (\cite{connes99}) was first to propose the use of square-section fiber to improve scrambling efficiency. Indeed, ray-tracing simulations show that such a section breaks the radial symmetry of fiber, which is responsible for
the radial scrambling inefficiency. But at that time, fiber manufacturers were unfortunately not 
able to build such a fiber. 

\section{SOPHIE radial velocity systematics due to inefficient scrambling of cylindrical fibers} 
\label{old}

\begin{figure}
   \centering
   \includegraphics[width=8cm]{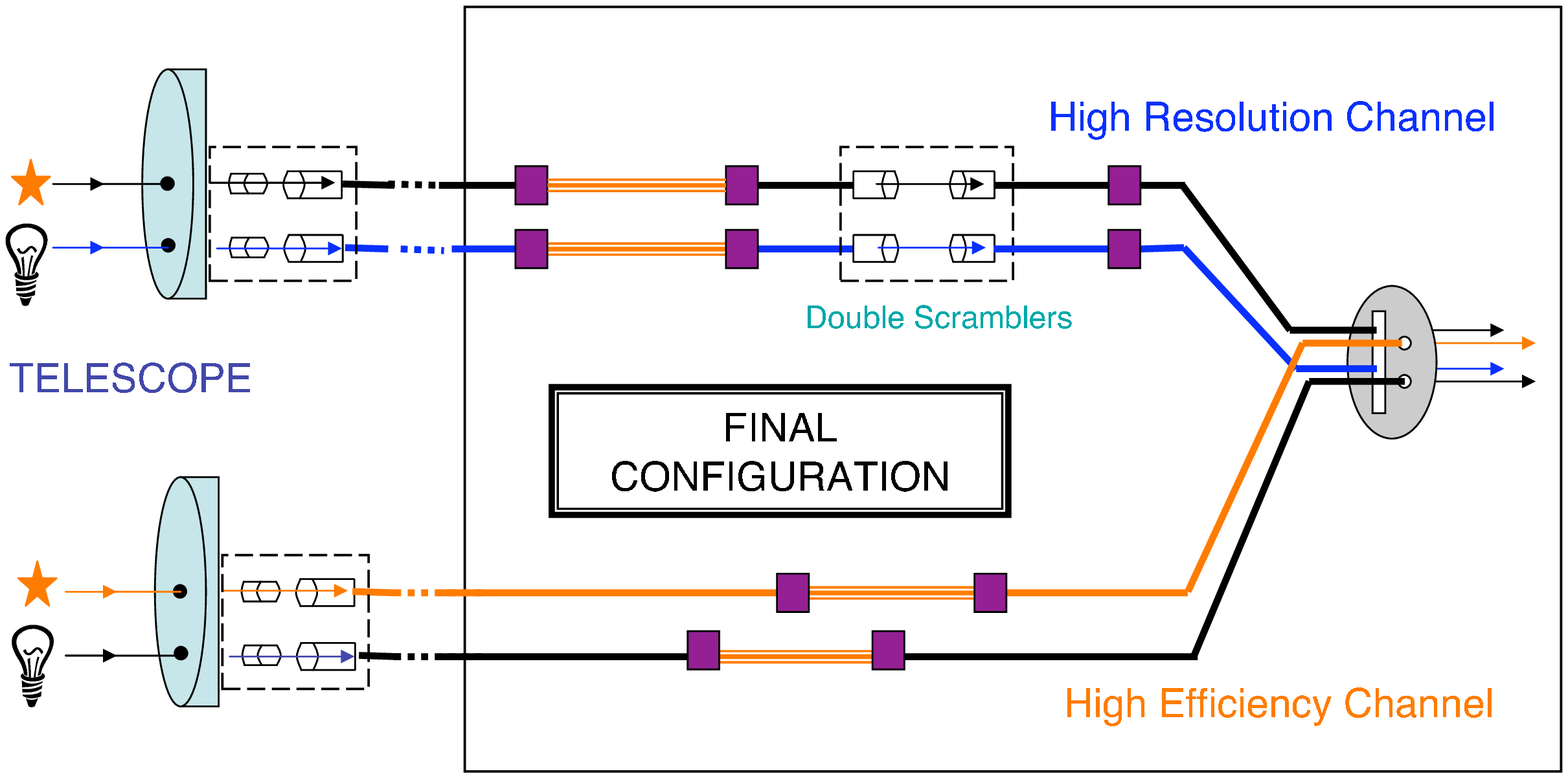}
      \caption{SOPHIE fiber links configuration. Octagonal fiber sections are shown as triple orange lines 
      between FC-FC connections, shown as purple squares.}
         \label{figfibres}
   \end{figure}

The SOPHIE spectrograph has two observing modes, both using standard step-index multi-mode cylindrical optical 
fibers (Polymicro FVP/100/110/125). In high-resolution (HR) mode, the spectrograph is fed by a 40.5 $\mu$m slit superimposed on the output of the 100 $\mu$m fiber, reaching a spectral resolution of  75,000. In high-efficiency mode (HE), the spectrograph is directly fed by the 100 $\mu$m fiber with a resolution power of 39,000. Both SOPHIE fibers have a sky acceptance of 3 arcsec, well adapted to median seeing condition at OHP. 
The high-resolution mode is equipped with a double-fiber scrambler. For each fiber pair, one aperture 
(fiber A) is used for starlight, whereas the other one (fiber B), 2 arcmin away from the first one, can be used either on a 
Thorium-Argon lamp for tracking spectrograph drift or on the sky to estimate background pollution, 
especially in case of moonlight. Figure~\ref{figfibres} 
shows the 
SOPHIE fiber links configuration.  

In order to quantify the effect of illumination and injection variations onto the fiber input 
to the RVs, several tests were performed on standard RV solar-type stars.

\subsection{Star decentering}

\begin{figure}
   \centering
    \includegraphics[width=9cm]{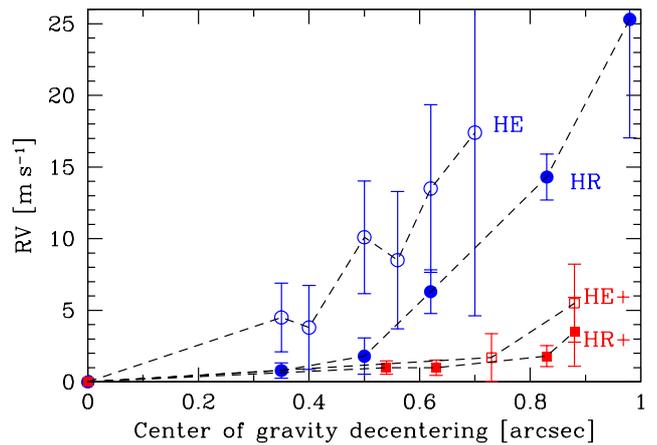}
     \caption{Star-decentering effect on SOPHIE radial velocities as function of the fiber illumination center of gravity for initial HE (blue open circle) and HR  (blue circle) modes and the new octagonal fiber links in HE+ (red open square) and HR+ (red square) modes.}
         \label{figdec}
   \end{figure}

{\modiff The tests of sensitivity to decentering effects were described in Perruchot et al. (\cite{perruchot11})}.
Figure \ref{figdec} (blue circle) shows the RV change as a function of the {\modiff fiber} illumination center of gravity for both HE and HR modes. The RV shift may reach up to 25 {\ms} for large decentering. These tests also show that HE mode has a sensitivity at least twice as larger as the HR mode and a strong sensitivity to the decentering offset direction illustrating the gain provided by the double scrambler. 
The typical accuracy of the SOPHIE new guiding system, implemented in September 2009, was estimated to be better than 0.3 arcsec. This corresponds to a smaller change in the illumination center of gravity (0.09 arcsec for a seeing of 3 arcsec). Thus, guiding and centering errors do not cause systematics larger than $\sim$1 {\ms} and $\sim$2 {\ms} for the HR and HE mode, respectively. 

Considering that the output fiber corresponds to a slit resolution of 7.7 {\kms} (resolution of 39,000), the scrambling gain, computed for a center of gravity displacement of 0.6 arcsec, is equal to 260 and 130 for the HR and HE mode, respectively. For comparison, the scrambling gain of HARPS fiber link was estimated to be 400. 
Furthermore, the HARPS spectral resolution (110,000) conducts to a 2.8 times smaller RV shift for a given illumination center of gravity offset. The SOPHIE HR mode is then at least four times more sensitive than HARPS to the 
injection and illumination change onto the fiber near-field.

\subsection{Telescope defocusing} 

\begin{figure}
   \centering
   \includegraphics[width=9cm]{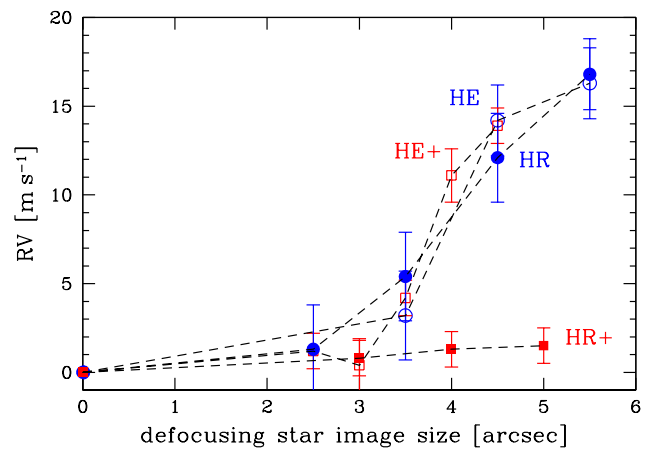}
     \caption{Star-defocusing effect on SOPHIE radial velocities for initial HE (blue open circle) and HR (blue circle) modes and for the new octagonal fiber links in HE+ (red open square) and HR+ (red square) modes.}
         \label{figfoc}
   \end{figure}

To test the sensitivity to telescope defocusing on the fiber entrance, we adjusted the telescope focus in order 
to enlarge the apparent size of the star image on the guiding camera. We explored different values from the seeing (2.5-3 arcsec), corresponding to the best focus, up to a FWHM image of 5.5 arcsec. 
Figure~\ref{figfoc} (blue circle) shows the RV change as a function of the apparent star image size introduced 
by the defocusing. Both HE and HR modes present about the same sensitivity with RV 
shift up to 15 {\ms} for large defocusing.  We noticed that for the largest defocusing, the shadow of the telescope secondary mirror started to be visible in the guiding camera and affected the guiding and centering system.

\subsection{Atmospheric dispersion}

\begin{figure}
   \centering
   \includegraphics[width=9cm]{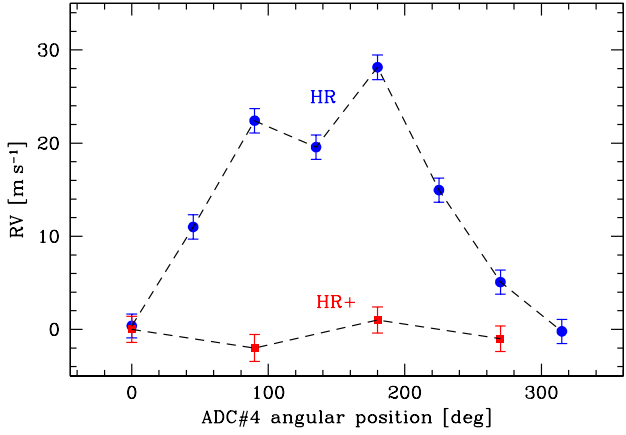}
     \caption{Atmospheric dispersion effect on SOPHIE radial velocities for initial HR (blue circle) mode and 
     for the new octagonal fiber links in HR+ (red square) mode.}
         \label{figadc}
   \end{figure}

The SOPHIE Cassegrain adapter contains an atmospheric dispersion correctors (ADC) to minimize as much as possible the loss of star light due to differential atmospheric refraction at the entrance of the fiber. This system is made of one parallel plate for a zenith angle smaller than 11 deg and of four normal field bi-prisms covering zenith angles 11-30, 30-44, 44-54, and 54-60 deg, respectively. The choice of the bi-prisms and their angles is set automatically by the observing system as a function of the airmass and the parallactic angle of the observed target. The residual of the atmospheric dispersion after correction is expected to never exceed 0.3 arcsec length up to an airmass of two.    
To test the effect of a wrong or incomplete correction of atmospheric dispersion on the fiber entrance, we adjusted manually the angle of the ADC. We performed this test at airmass close to 1.4 using the corrector number 4 
covering zenith angle 44-54 deg (airmass 1.39 - 1.70). For this corrector, the amplitude of dispersion from 387 to 694 nm provided by the bi-prism is 5 arcsec. By rotating the angle from its nominal value, 0 deg, up to 315 deg, we then expect to change the apparent size and orientation of the star image on the fiber entrance. Figure~\ref{figadc} (blue circle) shows the RV change in HR mode as a function of ADC angle. The enlargement of the image size, due to incomplete correction or even an increase of the dispersion (for 180 deg), leads to a RV shift up to 28 {\ms}.

\subsection{Seeing changes} 
\label{secseeing}

The seeing effect is not easy to test since we do not have the possibility to control it properly. Indeed, the defocusing 
of the telescope does not exactly reproduce a change of seeing. The pattern of the defocused star image is far from being like the seeing and furthermore defocusing affects both near- and far field on the fiber entrance. 
The seeing effect was identified on the HR mode when we found a correlation between the signal-to-noise ratio (SNR) and the radial velocities on several standard stars observed with the same exposure time. 
This effect, first described by Boisse et al. (\cite{boisse10a}, \cite{boisse10b}), appeared to be the larger limiting factor in the SOPHIE precision. Whereas all the effects described above (star decentering, telescope defocusing, atmospheric dispersion) may be controlled, this is not the case for the seeing. 

The seeing value is not automatically monitored for each SOPHIE observation. To roughly estimate the 
seeing at the fiber entrance, we compute the SNR per pixel 
at 550 nm and assume that this SNR only depends on the exposure times $T_{exp}$, target 
magnitude $mv$ and the fiber filling factor $F_f$. The filling factor $F_f$ corresponds to the transmission ratio of a 
two-dimensional Gaussian profile with a FWHM equal to the seeing on a circular fiber of 3-arcsec diameter. 
Neglecting detector read-out noise and atmospheric absorption, the number of photon-electrons 
per pixel at 550 nm is given by\\

\begin{equation}
N_{phot} = SNR^2 =  \frac{ N_0 . T_{exp}  .  F_f }{ 2.512^{mv} }  
\label{equa1}
\end{equation}
with 
\begin{equation}
F_f = 1 - exp(-0.684.(\frac{3}{seeing})^2 \,.
\end{equation}

N$_0$ corresponds to the expected number of photon-electrons at 550 nm per pixel and per second 
for mv=0. This factor, which was estimated to 5.3 10$^4$, includes the atmosphere, the telescope, the fiber link, 
and the instrumental efficiency at 550 nm. We checked that some seeing values derived from Eq.~\ref{equa1} 
were in agreement with the corresponding values measured on the guiding camera. The seeing value is a rough estimation and an upper limit of the true seeing because atmospheric absorption cannot be estimated.  

   \begin{figure}
   \centering
   \includegraphics[width=9cm]{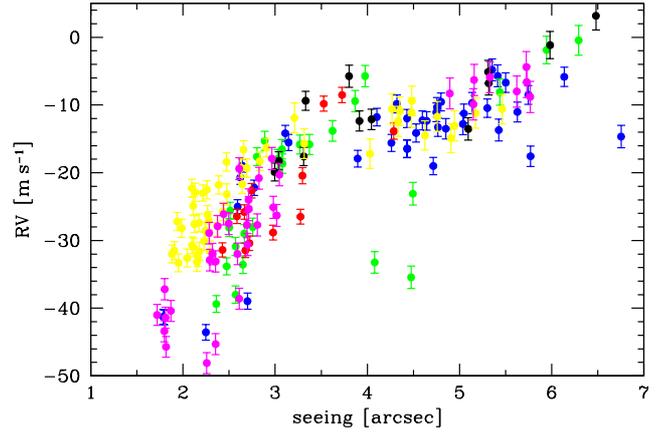}
     \caption{SOPHIE radial velocities of standard stars for initial HR mode as function of the seeing. Each of the six standard stars is represented by a different color (see Table~\ref{tableHR}).}
         \label{figseeing}
   \end{figure}

Figure \ref{figseeing} shows the RV of six standard stars measured with the HR mode as a function of the seeing estimation. These stars are listed in Table~\ref{tableHR}. To adjust the velocity offset, we removed the averaged RV of each star measured later with the new octagonal fiber link (see section \ref{sec42}). 
We see a clear correlation between the seeing and the RV. The seeing variation may introduce RV change 
of up to 35 {\ms} peak to peak in the worst case. Some measurements below the main curve probably correspond to measurements made under bad weather conditions, with atmospheric absorption due to cirrus, which leads to an overestimate of the seeing. 
Two regimes seem to appear on Fig.~\ref{figseeing} :  1) for seeing larger than the fiber aperture 
(3 arcsec), the correlation between RV and seeing is small, if not negligible (3.6$\pm$1.9 {\ms  arcsec$^{-1}$}) ;  2) for seeing smaller than the fiber aperture, the correlation becomes stronger and very significant (22.2$\pm$8.9 {\ms arcsec$^{-1}$}). In this regime, when the seeing is improving (decreasing), mainly the center of the fiber entrance is illuminated. Taking into account the double scrambler, the far field at the fiber output is then mainly illuminated on its central part. As shown in Boisse et al. (\cite{boisse10a}), ray-tracing simulation demonstrates that a center-illuminated far field projected on the SOPHIE grating induces change of RV that are not symmetric along the spectral orders 
and then introduces an apparent change in RV. The seeing variations on the fiber entrance 
are directly projected on the spectrograph grating. The grating is then more or less center-illuminated 
with a direct effect on the RV shift. We did not find similar correlation with the HE mode, which confirms that 
seeing variations are not comparable to telescope defocusing.

   \begin{figure}
   \centering
   \includegraphics[width=9cm]{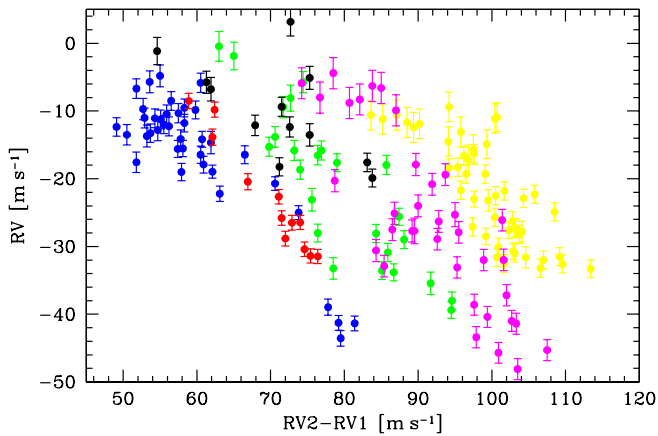}
     \caption{SOPHIE radial velocities of standard stars for initial HR mode as function of the radial velocity 
     difference $RV2-RV1$ of right and left sides of spectral orders. Each of the six standard stars is represented by a 
     different color (see Table~\ref{tableHR}).}
         \label{figvrrb}
   \end{figure}

To check the RV change along spectral orders, we cut each spectral order in its middle and 
computed the RV on each side. We call $RV1$ the RV computed using only the left (or blue side) 
of spectral orders and $RV2$ the RV derived from their right side (or red side). The difference $RV2-RV1$ 
is not equal to zero but strongly correlated with the seeing. 
Figure \ref{figvrrb} shows the RV of the same six standard stars as in Fig.~\ref{figseeing} 
but plotted as function of $RV2-RV1$.  A clear anti-correlation appears between RV and the difference $RV2-RV1$, with a slope of -1.04$\pm$0.27, illustrating the strong sensitivity of the SOPHIE spectrograph to pupil illumination change. The average value of $RV2-RV1$ appears to be function of the $B-V$ index. 

The RV dispersion of the six standard stars shown in Fig.~\ref{figseeing} and Fig.~\ref{figvrrb} are listed in Table~\ref{tableHR}. They range from 7.1 to 12.6 {\ms}.  The seeing effect could be partially corrected using the correlation between RV and seeing (as done by Boisse et al. \cite{boisse10b}, \cite{boisse12}) or the correlation between RV and the difference $RV2-RV1$ (as done by Diaz et al. \cite{diaz12}). 

We found that in the HR mode all effects that introduced an increase of the star image in the fiber entrance 
(decentering, defocusing, atmospheric dispersion, and seeing) lead to an apparent increase of the RV.

\subsection{Mechanical and thermal change on the fiber links}
\label{secmecha}

In the previous sections, we emphasize the fact that SOPHIE radial velocities  
are extremely sensitive to the injection conditions and illumination of the fiber input. 
We also made some tests consisting of slightly displacing and bending the fiber link between the 
Cassegrain unit and the Coud\'e train. We saw some effect at the level of few {\ms}, 
indicating that bends inside the fiber may slightly change the illumination. 
Although it was difficult to test, we have some hints of evidence that the temperature 
inside the dome (similar to outside temperature) may also impact on the fiber illumination. 
The correlation coefficient found between RV and the difference $RV2-RV1$ seems to be a function 
of the epoch of observations along the year, indicating that the temperature of the fiber 
(inside the dome) may slightly affect its scrambling properties. Some RV changes observed in the HE mode 
seem to be related to temperature change inside the dome, which may affect the far field and then 
introduce grating illumination variations.

\section{Laboratory and on-sky properties of octagonal fibers} 
\label{sec4}

\subsection{Octagonal fiber characteristics}

Considering the low scrambling efficiency of the SOPHIE fiber link and the strong sensitivity of the spectrograph 
to illumination change, we decided to implement octagonal fibers on the fiber links. The octagonal section was chosen 
in order to have a better coupling efficiency or filling factor (94.8\%) with circular-section fibers in comparison with square-section (78.5\%). 

   \begin{figure}
   \centering
   \includegraphics[width=4.55cm]{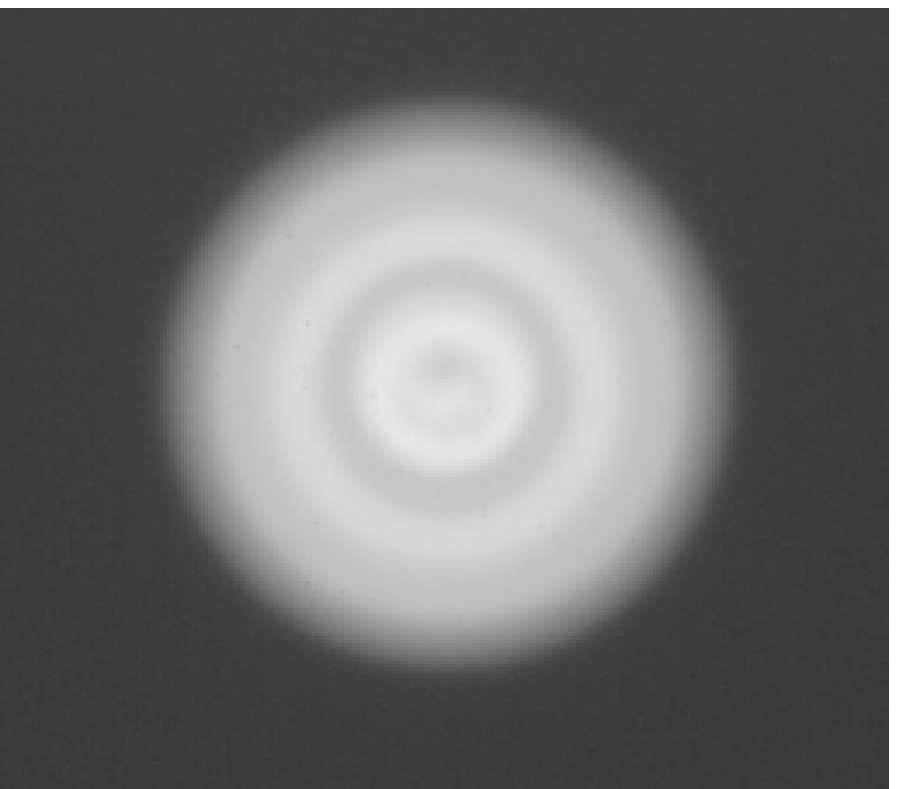}
   \includegraphics[width=4.23cm]{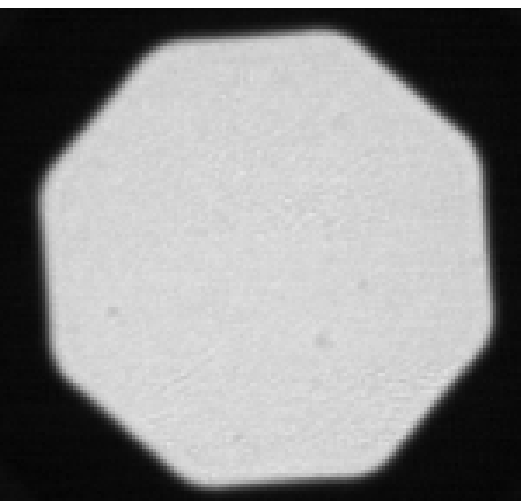}
     \caption{Pictures of the section of the 100 $\mu$m diameter SOPHIE circular-section fiber (left) and the octagonal-section fiber (right).}
         \label{figphoto}
   \end{figure}

The scrambling properties were measured using the test bench of the Geneva Observatory described 
in Chazelas et al. (\cite{chazelas10}). 
Figure~\ref{figphoto} shows a picture of the face of the initial SOPHIE fiber and the new octagonal-section fiber. 
{\modif Our laboratory test, described in Perruchot et al. (\cite{perruchot11}), led to} a scrambling gain of 180 for the SOPHIE-type circular fiber, in agreement with our tests on sky made with the HE mode (130) and greater than 3500 for the octagonal one. 

On 2011 June 17, 1.5 m length octagonal fibers were inserted on fiber A and B of the HR mode in front of the double scrambler and on fiber A of the HE mode. On 2011 October 10, the HE mode was completed with the insertion of a 1.5 m length octagonal fiber on fiber B. The technical details about the octagonal fiber, its implementation on the SOPHIE fiber link, and first tests are presented by Perruchot et al. (\cite{perruchot11}). Figure~\ref{figfibres} 
shows the new SOPHIE fiber link configuration, including the octagonal fibers as in 2011 October 10. 
The flux loss due to the insertion of octagonal fiber was estimated to be 8\% and 10\% on fiber A of HR and HE mode, 
respectively. This efficiency loss in the science fibers is as expected due to the coupling efficiency 
of an octagonal section with circular section (5.2\%) and the FC-FC connection loss (2-5\%) measured in 
the laboratory. The SOPHIE spectrograph with these new octagonal fibers was renamed SOPHIE+.
With such an implementation, we expect a very high scrambling of the near-field illumination of the fibers, which is projected for the HR mode in the spectrograph pupil. We do not expect a gain in the scrambling of the far field. 

\begin{table*}
\caption{Radial velocity standard solar-type stars measured on HR mode with SOPHIE and SOPHIE+}             
\label{tableHR}      
\centering                         
\begin{tabular}{llll|ccc|ccc|c}        
\hline\hline                 
\multicolumn{4}{c}{Target}  &  \multicolumn{3}{|c|}{SOPHIE\_HR} & \multicolumn{3}{|c|}{SOPHIE\_HR+} & SOPHIE\_HR+ \\
Name & spectral & mv  &  Fig. color & $\sigma_{RV}$  & span & N$_{obs}$ & $\sigma_{RV}$  & span & N$_{obs}$ &  10 nights $\sigma_{RV}$ \\    
& type & & & [{\ms}] & [days] & & [{\ms}] & [days] & & [{\ms}] \\
\hline                        
HD\,109358  & G0V  & 4.3  & yellow  & 7.6  &  97  & 48  & 2.4  & 21  & 80  & 2.0   \\
HD\,139324  & G5IV-V & 7.5 & green &  11.1  &  93  & 25  & 2.2  & 31  &  11 & 2.2 \\ 
HD\,147512 &  G0V  &  7.3 & black  & 7.1  &  79  & 12  & - &  - & - & - \\
HD\,185144  & G9V   & 4.7  & red  & 7.8  & 40  & 12  &  1.5 & 97  & 63  & 0.9 \\
HD\,30708  &  G5V & 6.8  & blue  & 9.0  &  360 & 39  &  2.3  & 92  & 13  & 1.2 \\
HD\,55575  & G0V  & 5.6  & magenta & 12.6  & 16  & 36  & 3.4  & 118  & 56  & 1.5 \\
HD\,221354  & K2V   &  6.7 & black &  - & - & - & 2.2 & 174 & 43 & 0.9 \\
HD\,9407  &  G6V  &  6.5  &  cyan  &  - & - & - & 2.7  &  160  & 23  & 1.6 \\
\hline                                   
\end{tabular}
\end{table*}

\subsection{Tests of systematic effect on standard stars}
\label{sec42}

To quantify the gain of scrambling provided by the insertion of an octagonal fiber and to estimate the precision of SOPHIE+ under real observing conditions, we performed the same tests as those described in Section \ref{old} on 
standard solar-type stars.

The tests of decentering (see Fig.~\ref{figdec}) show that the sensitivy to the guiding decentering is reduced 
by at least a factor of six in both HR+ and HE+ modes illustrating the gain of scrambling in the near field. 
The scrambling gain, computed for a center of gravity displacement of 0.6 arcsec, is equal 
to $\sim$1500 and $\sim$1000 for the HR and HE mode, respectively. The typical accuracy of the new guiding system, 0.3 arcsec, is not expected to introduce RV jitter larger than 0.5 {\ms}.  

The tests of defocusing (see Fig.~\ref{figfoc}) show that the sensitivity to the telescope defocusing 
is not reduced on the HE+ mode but reduced by a least a factor 6 for the HR+ mode. The non improvement 
in the HE+ mode may indicate that defocusing affect both the near- and far field, the last-one being 
not scrambled by the octagonal fiber. The tests of a wrong or incomplete correction of the atmospheric 
dispersion (see Fig.~\ref{figadc}) show that in HR+ mode there is a real improvement with no more 
significant effect on the RV.  

\begin{figure}
   \centering
   \includegraphics[width=9cm]{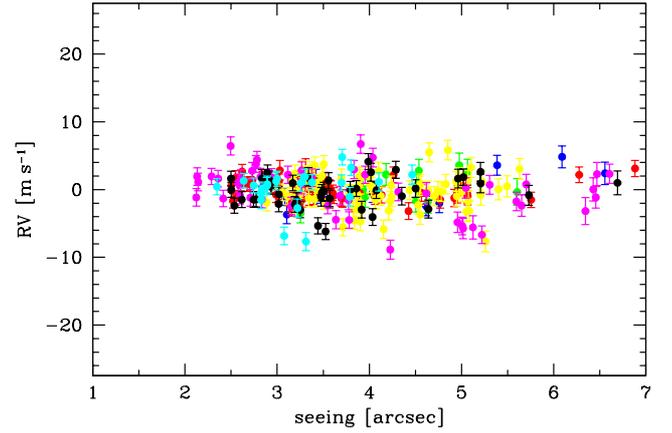}
     \caption{SOPHIE+ radial velocities of standard stars observed with the final HR+ mode, including octagonal fibers, as function of the seeing. This figure should be directly compared with Fig.~\ref{figseeing}. Each of the seven standard stars are represented by a different color (see Table~\ref{tableHR}).}
         \label{figseeing2}
   \end{figure}

To estimate the seeing effect, we monitored within the HR+ mode seven standard stars 
as in section \ref{secseeing}. These stars are listed in Table~\ref{tableHR}.
The RV of these standard stars are plotted in Fig.~\ref{figseeing2} 
as a function of the seeing value. No more correlations are seen between RV and seeing. 
The RV dispersions of the seven standard stars shown in Fig.~\ref{figseeing2} are listed in Table~\ref{tableHR}. 
They range from 1.5 to 3.4 {\ms}, indicating a gain of about five in the RV precision for the HR+ mode. Three of these targets were monitored with HIRES on the 10.2 m Keck telescope: HD185144 (Howard \cite{howard10}), HD221354, and HD9407 (Howard \cite{howard11}), with RV dispersion of 2.0, 1.9, and 1.7 {\ms} respectively over several years.  
For the HR+ mode, we also report in Table~\ref{tableHR} the dispersion obtained over a time scale 
of ten nights. During this time span, the RV dispersion ranges from 0.9 to 2.2 {\ms}, very close to the 
photon noise uncertainty. To explain the slightly worse precision during time spans longer than ten nights, we suspect possible remaining systematic effects. The thermo-mechanical evolution of the far field in the fiber link of  the HR+ mode is not scrambled by the octagonal fiber. This far field is converted by the double scrambler to the near field at the entrance of the spectrograph and may introduce a long-term effect at the level of few {\ms}. 
If this is confirmed, we will have to consider inserting an additional octagonal fiber after the double scramblers 
(see Fig.~\ref{figfibres}). The evolution of the Thorium-Argon lamp, 
{\modif with a life span of about one year on SOPHIE},  
and more precisely the flux ratio between Thorium and Argon lines, {\modif which may evolve on a time scale of few weeks}, are presently not taken into account in the data-reduction pipeline. Finally, we cannot exclude for some of our targets intrinsic variability on the long term at the level of few {\ms}. 

In Table~\ref{tableHE}, we list the RV dispersion measured on standard stars with the initital HE mode 
and the new HE+ mode.  For the HE+ mode, the RV precision, between 3 and 4 {\ms}, is not as good as the HR+ mode. This may indicate that the far field in HE+ link, which is not scrambled by the octagonal fibers, 
is not sufficiently stable. As discussed in section~\ref{secmecha}, we suspect that this far field is sensitive to the thermomechanical environment of the HE+ fiber link. There is presently no way to improve this mode. However, the high-efficiency mode is not devoted to the high-precision RV measurements and is usually used in 
photon-noise limited regimes for targets fainter than mv=12.

\begin{table}
\caption{Radial velocity standard solar-type stars measured on HE mode with SOPHIE and SOPHIE+}             
\label{tableHE}      
\centering                         
\begin{tabular}{c|ccc|ccc}        
\hline\hline                 
Target  &  \multicolumn{3}{|c|}{SOPHIE\_HE} & \multicolumn{3}{|c}{SOPHIE\_HE+} \\
Name  &   $\sigma_{RV}$  & span & N$_{obs}$ &   $\sigma_{RV}$  & span & N$_{obs}$\\    
            & [{\ms}] & [days] & & [{\ms}] & [days] &  \\
\hline                        
HD\,109358  & 10.9  &  103  & 42  & -  & -  & -     \\
HD\,139324  &  25.0  &  93  & 34  & 3.8  & 87  &  7  \\ 
HD\,147512  & 22.7  &  83  & 15  & - &  - & - \\
HD\,185144  & -  & -  & -  &  2.2 & 243  & 25   \\
HD\,30708    & 6.0 &  16 & 36  &  3.7  & 92   & 15    \\
HD\,55575    & 9.0  & 21  & 40  & -  & -  & -   \\
HD\,221354  &  - & - & - & 3.1 & 91  &  32   \\
HD\,9407       &  - & - & - & 3.6  &  160 & 27  \\
\hline                                   
\end{tabular}
\end{table}

\subsection{Internal accuracy}

{\modiff As described in Perruchot et al. (\cite{perruchot11}), the drift differences measured on double Thorium-Argon calibration and reported in Table~\ref{tabledrift} are strongly reduced by the use of octagonal fibers.}
In the previous configuration, relative displacement of the fibers A and B as well as thermomechanical changes in the Cassegrain adapter led to relative change of illumination in fiber pairs. With the new configuration, the relative drift in HR+ mode is at the level of 22 {\cms} RMS to be compared to the photon-noise of Th-Ar spectra of 10 {\cms}. Even with the calibration lamp, which provides stable illumination of the fiber entrance, we find an improvement using octagonal fibers.\\

All our different tests show that the new octagonal fiber link provides a real and significant 
improvement in RV stability, especially with the HR+ mode. 
The scrambling gain (1000 and 1500 for HE+ and HR+ mode, respectively), if not as impressive as 
in the laboratory (3500), increased by a factor $\sim$6.

\begin{table}
\caption{Relative drift between fiber A and B measured with the Thorium-Argon lamp}             
\label{tabledrift}      
\centering                         
\begin{tabular}{ccccc}        
\hline\hline                 
Mode & HE   &  HE+   &  HR   &  HR+     \\    
\hline
Relative drift [{\ms}] &  2.27 & 0.58   &  0.57   &  0.22  \\
\hline                                   
\end{tabular}
\end{table}

\section{Scientific validation of SOPHIE+}

In order to validate the new performances of SOPHIE+ with octagonal fibers for Doppler asteroseismology and exoplanetology, we made RV series on specific solar-type stars. 

\subsection{Doppler seismology of the Sun and solar-type stars}

To validate the capability of SOPHIE+ to detect the tiny acoustic oscillations modes (p-modes) 
of solar-like stars, we performed high-cadence RV 
sequences of several hours as in Doppler asteroseismology runs. We observed the Sun, HD139324, Procyon, 
and Eta Cas. For each target, we computed the RV dispersion $\sigma_{RV}$, the noise level in the amplitude spectrum $\sigma_{TF}$ in a frequency domain free of oscillation modes, and the corresponding noise in the time series $\sigma_{noise}$ computed as $\sigma_{TF} . \sqrt{N_{obs} / \pi}$ with $N_{obs}$ the number of RV measurements. The results are summarized in Table~\ref{tableastero} and plotted in Fig.~\ref{figastero}.

\subsubsection{Helioseismology}

To measure the p-mode oscillations of the Sun, we oriented the telescope to blue sky during day time and made a 3 h sequence totaling 200 observations. The exposure time was set to 20 sec and the dead-time between exposures 
was typically 34 sec. The HR+ mode with simultaneous Thorium-Argon was used. The SNR {\modif at 550 nm} ranges from 270 to 310, {\modiff and the computed} RV photon noise\footnote{\modif{Blue-sky spectra have much more signal in the blue spectral orders than solar-type spectra for a given SNR at 550 nm. The photon noise is then accordingly smaller {\modif than for a solar-type star observed at the same SNR at 550 nm}.}} {\modiff is between} 35-40 {\cms}.

\begin{figure*}
   \centering
    \includegraphics[width=8cm]{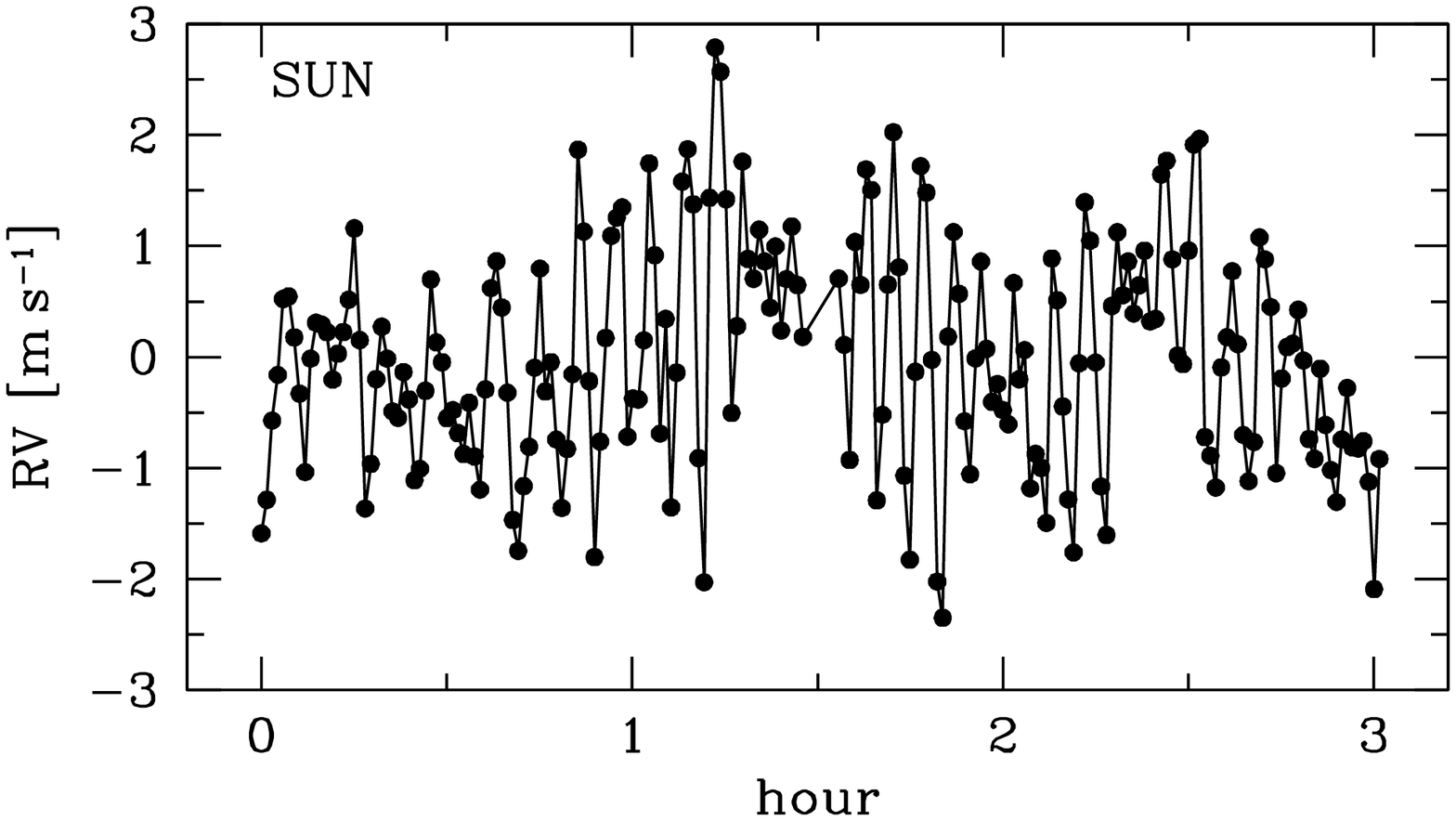}
     \includegraphics[width=8cm]{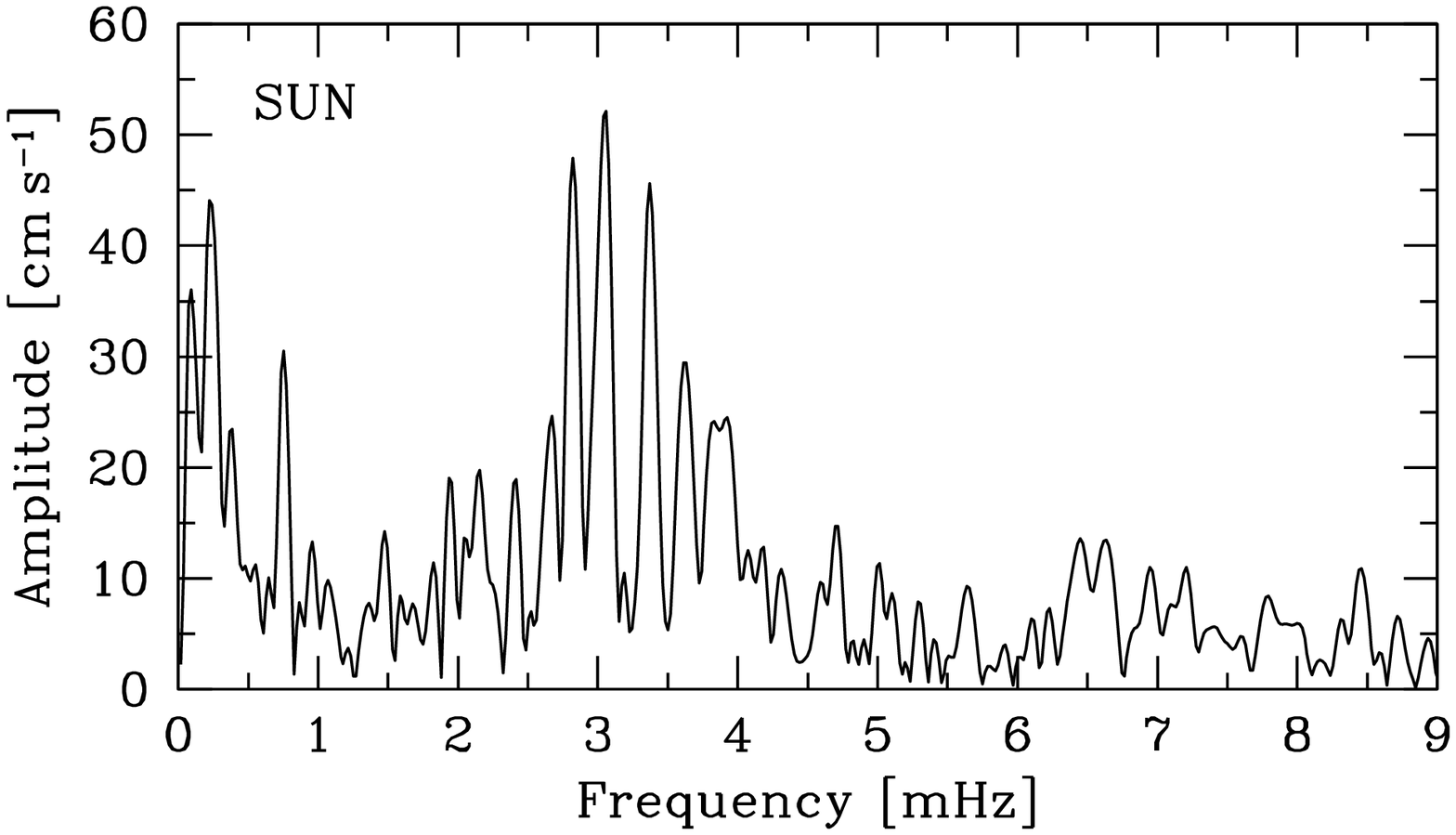}
    \includegraphics[width=8cm]{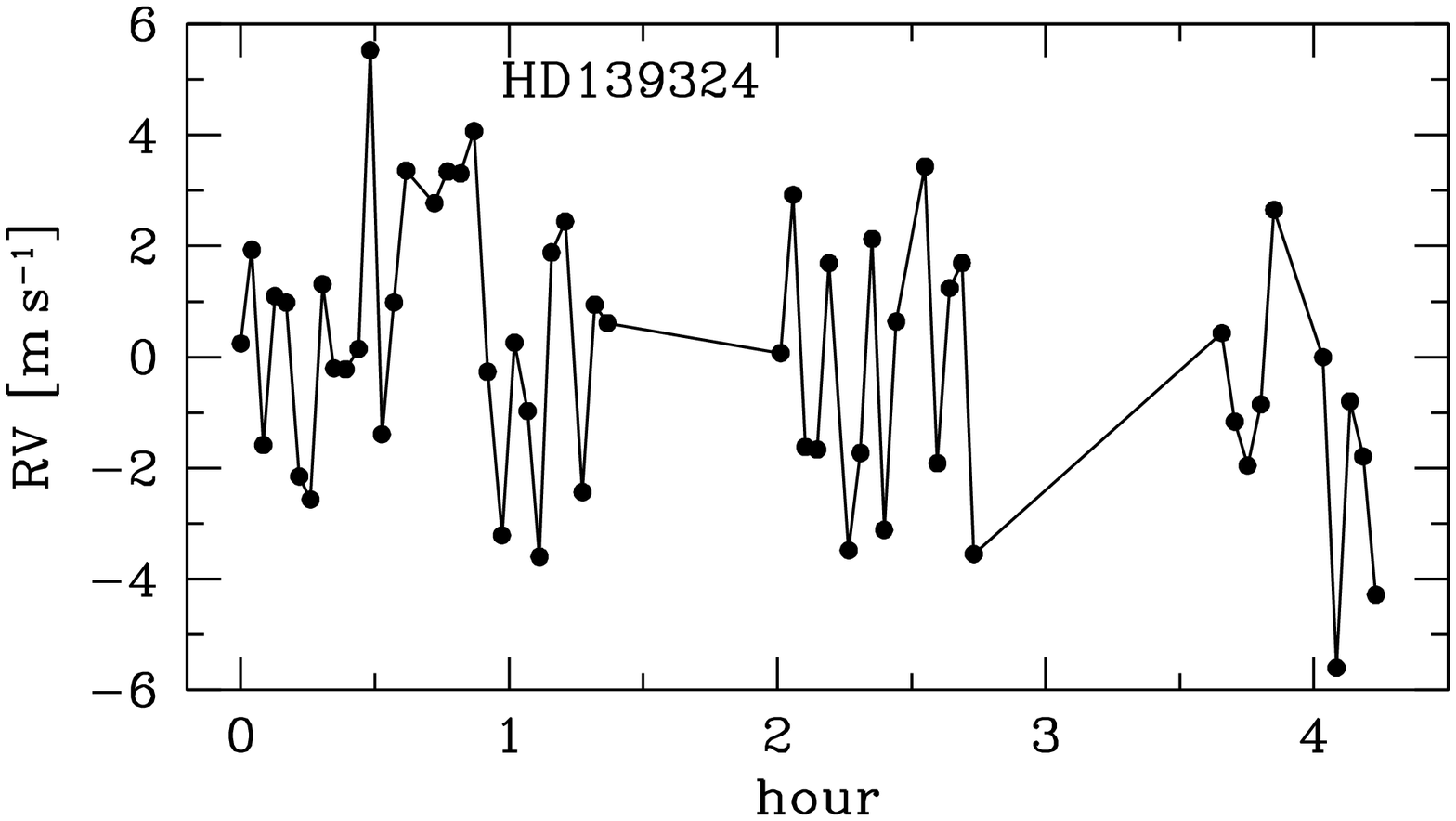}
     \includegraphics[width=8cm]{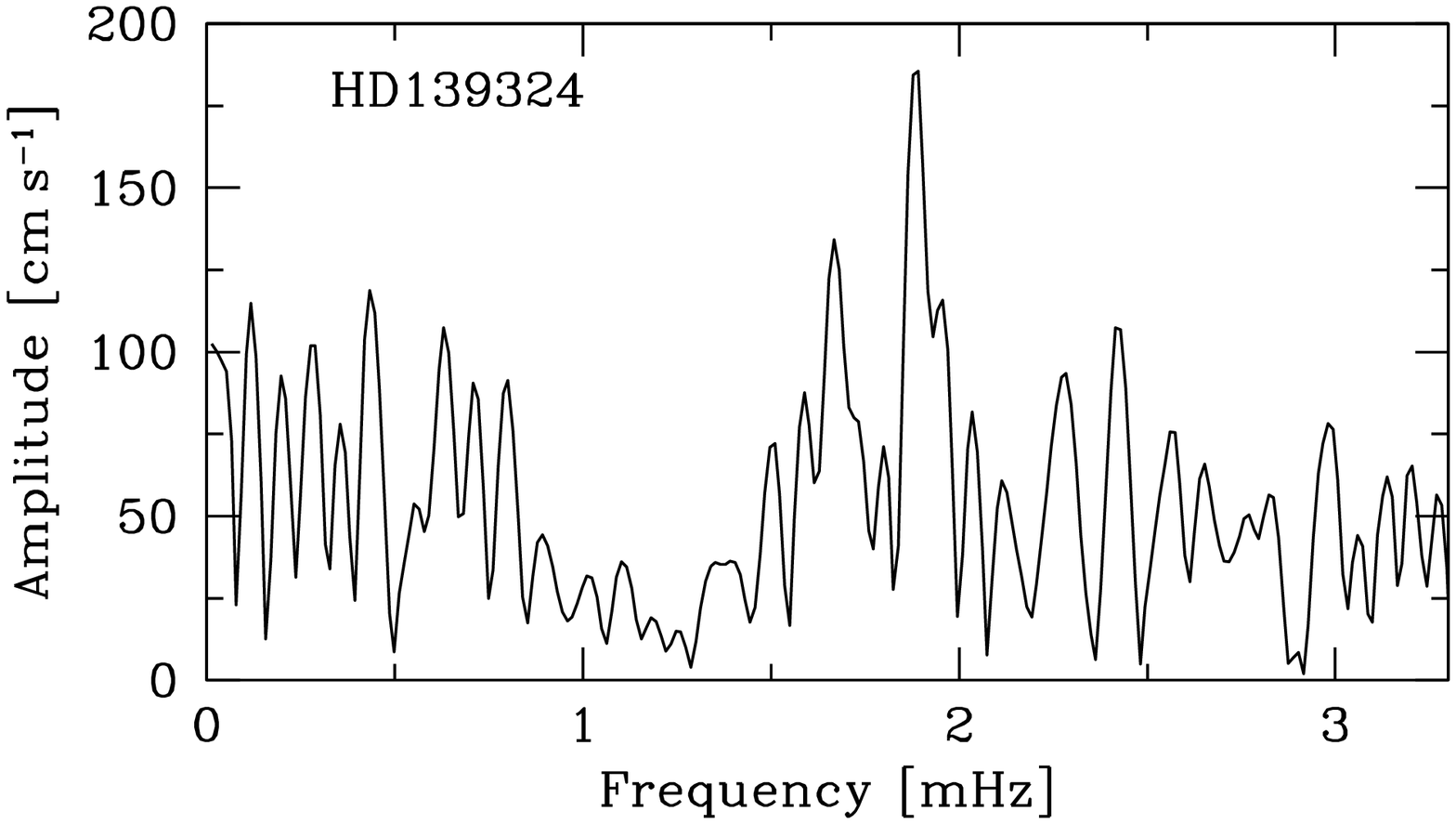}
    \includegraphics[width=8cm]{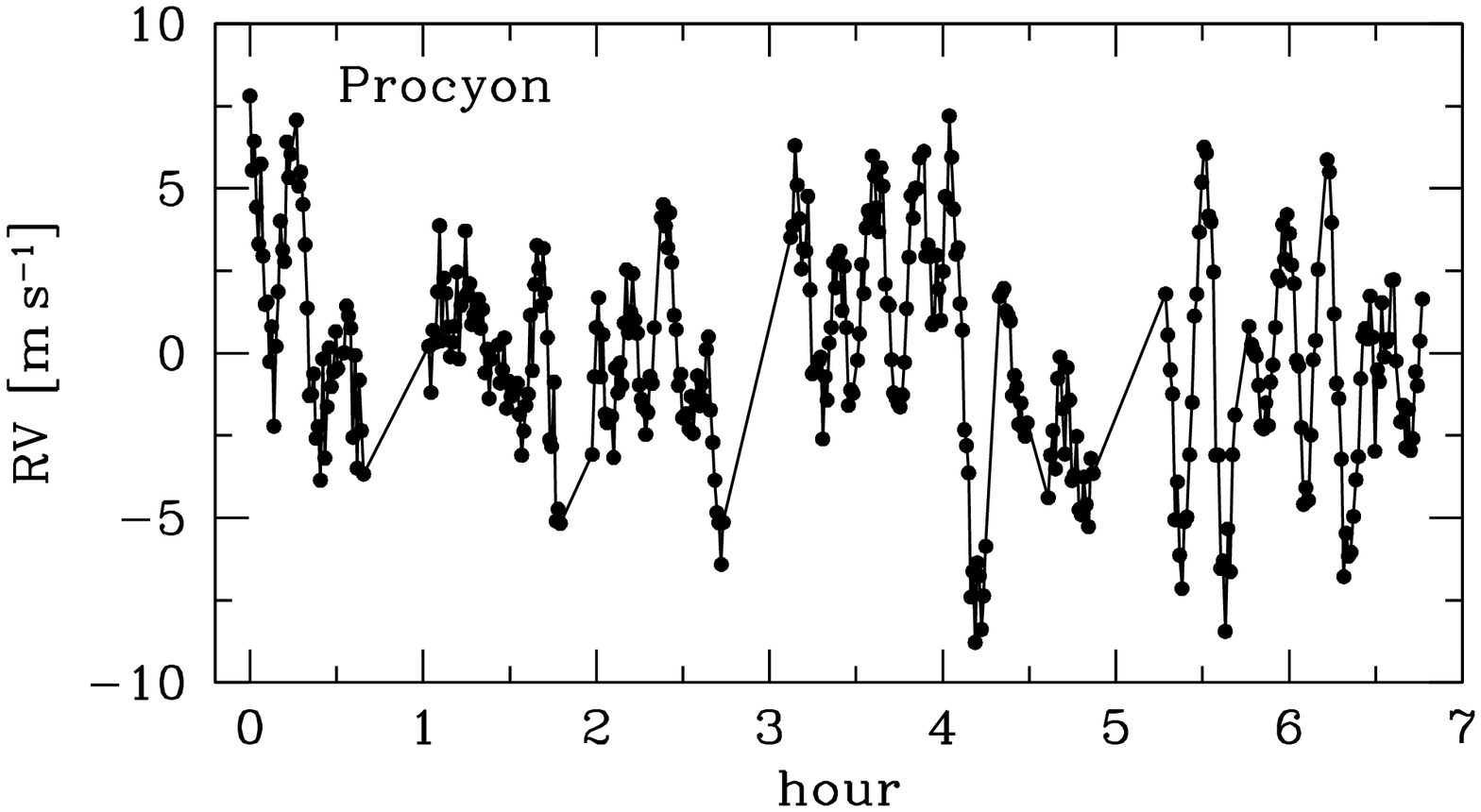}
     \includegraphics[width=8cm]{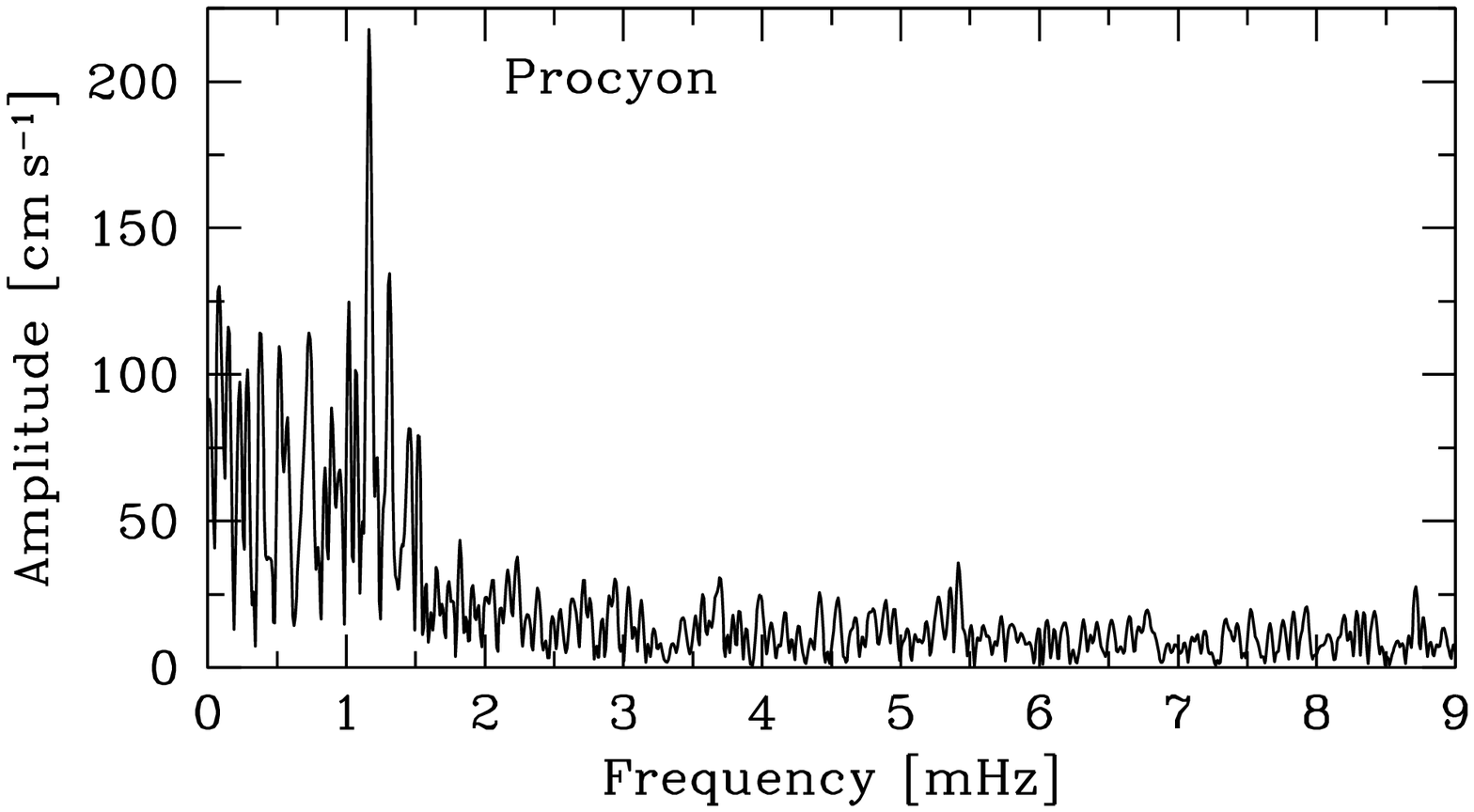}
    \includegraphics[width=8cm]{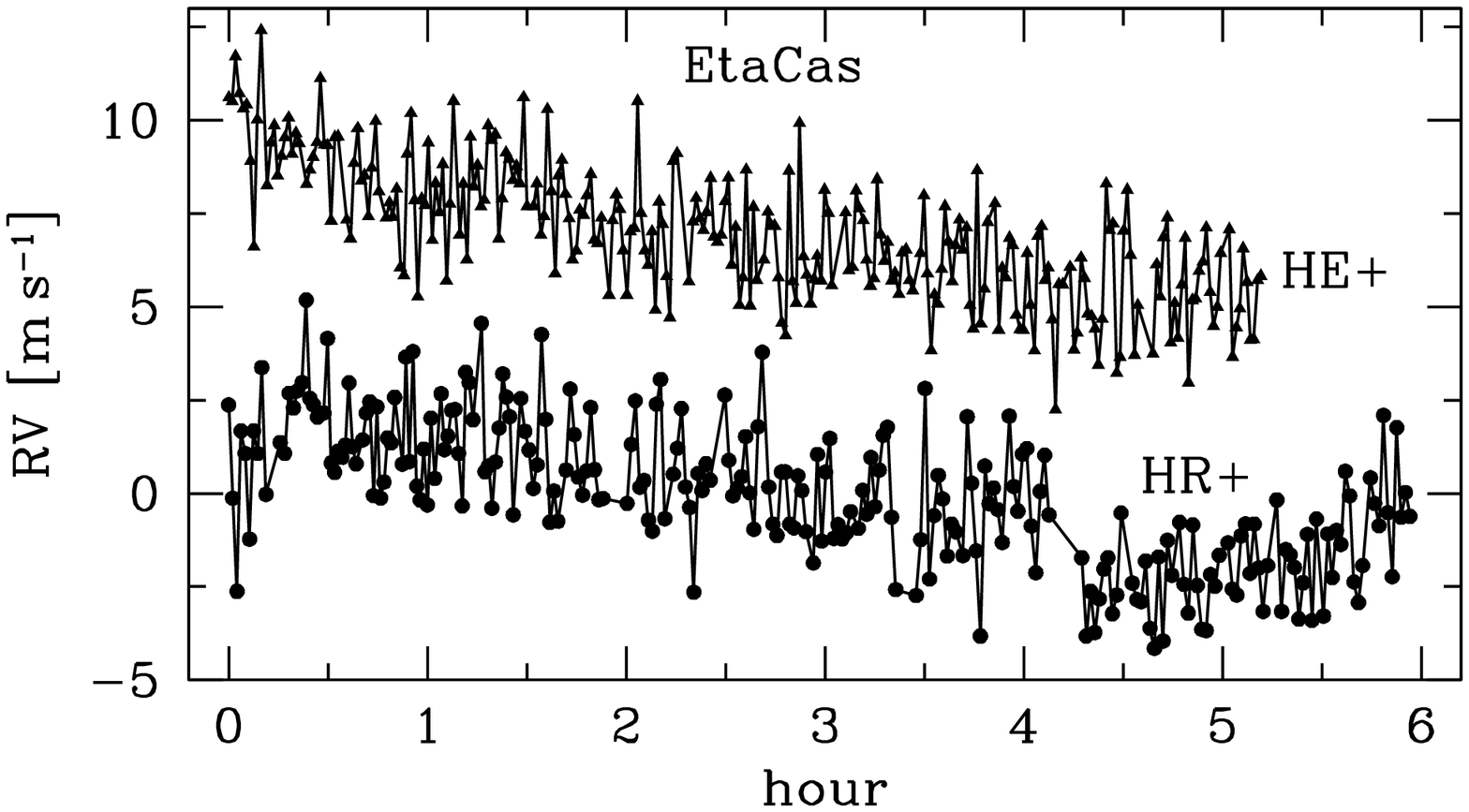}
     \includegraphics[width=8cm]{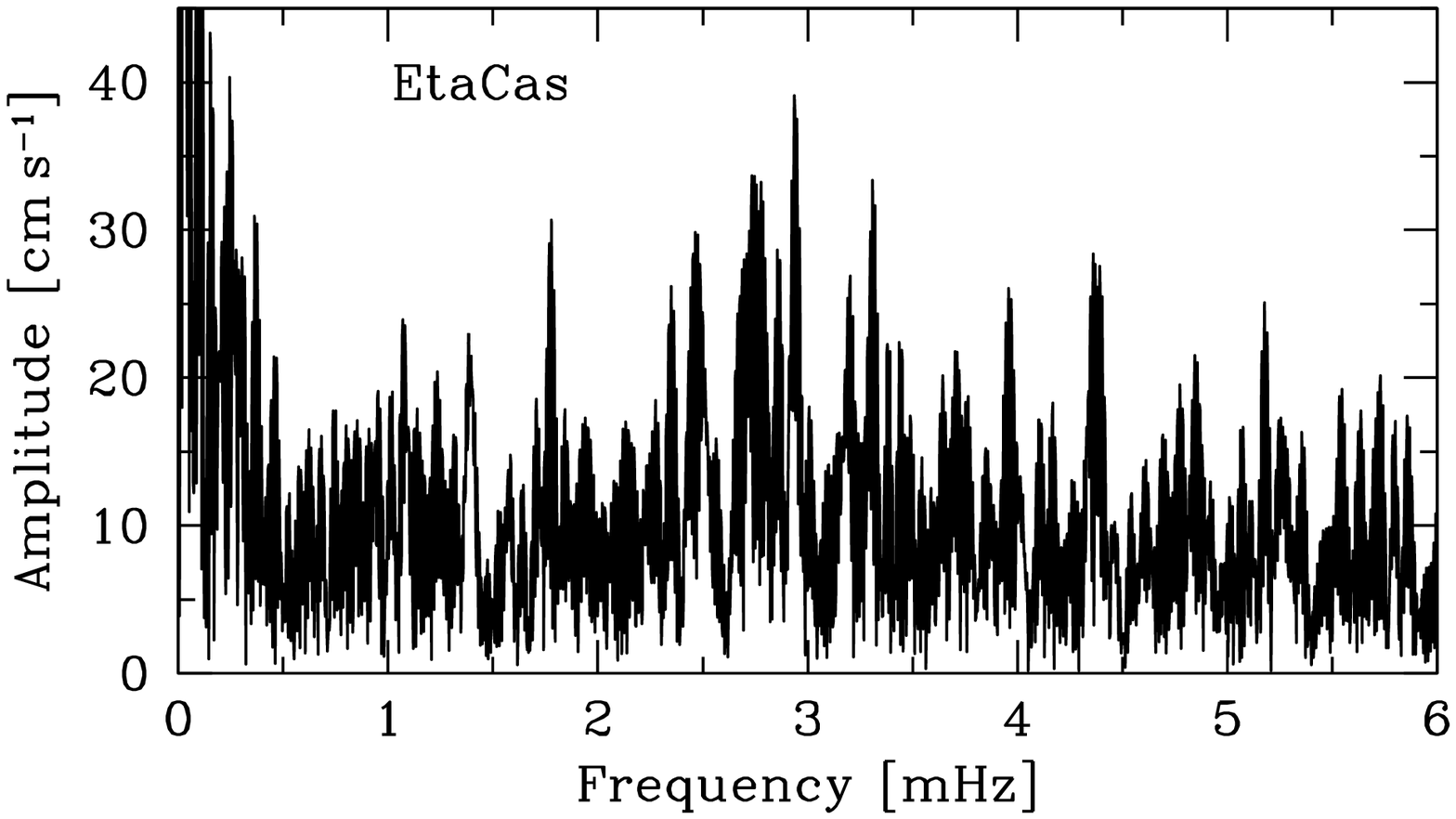}
    \caption{Radial velocity sequences obtained with SOPHIE+ on the Sun and solar-type stars. Left panels show the radial velocity time series. Right panels show the amplitude spectrum.}
         \label{figastero}
   \end{figure*}

A linear drift of 7.5 {\ms}  was observed in our sequence, which may be related to the 
inappropriate barycentric Earth RV due to wrong sun coordinates and/or the effect of 
diffused atmosphere like winds, which were not taken into account. 
We then subtracted a slope of 2.5 {\ms}hour$^{-1}$. Figure~\ref{figastero} shows the RV obtained 
on the blue sky after removing the slope. The dispersion is equal to 1.0 {\ms} but is clearly dominated 
by the oscillation modes with period close to 5 mn. Averaging ten consecutive data points  
led to a dispersion of 47 {\cms}. Figure~\ref{figastero} shows the amplitude spectrum of the 
RV. The p-modes are clearly visible with their comb-like structure around 3 mHz (corresponding 
to $\sim$5 minutes). A mean noise level of 5.4 {\cms}  was computed in  the amplitude spectrum in the frequency range 
5.5 - 9 mHz. Considering our 200 RV measurements, the corresponding velocity noise in the 
time series is 43 {\cms}. Taking into account the Sun photon noise of 37 {\cms} and the Thorium-Argon 
photon noise of 16 {\cms}, the expected noise is 40 {\cms}, which is very close to the measured noise 
at high frequency. 
We note that such a test on the blue sky is totally free of any illumination variation and may 
be representative of the present ultimate limit of SOPHIE+, which intrinsically can reach precision 
well below 1 {\ms} on short time scales.

\subsubsection{Asteroseismology of HD139324}

A sequence of RV measurement was made on the G5IV-V star HD139324 using the HE+ mode 
of SOPHIE+ without simultaneous Thorium-Argon. The exposure time was adjusted to between 120 to 150 sec 
in order to have an almost constant SNR equal to 95 (thanks to the exposure meter) {\modiff and a computed} RV photon noise of 172 {\cms}. The sequence of 54 measurements spread over 
4.3 hours was interrupted twice for Thorium-Argon calibration. We used these calibration to determine the spectrograph drift, which was typically between 1 and 3 {\ms}hour$^{-1}$, and interpolated it to the dates of our observations in order to correct them. Figure~\ref{figastero} shows the radial velocities of HD139324. The dispersion is 
equal to 2.35 {\ms} but one may see in the RV series that oscillation modes introduce some signal. 
Figure~\ref{figastero} shows the amplitude spectrum of the radial velocities. The p-modes are visible around 2 mHz 
(corresponding to $\sim$8 minutes). A mean noise level of 44 {\cms}  was computed in the amplitude spectrum in the frequency range 2.5 - 3.3 mHz. Considering our 54 measurement, the corresponding velocity noise in the 
time series is 182 {\cms}, which is very close to the estimated photon noise.

\subsubsection{Asteroseismology of Procyon}

A sequence of RV measurements was made on the well-known asteroseismological target Procyon 
using the HR+ mode with simultaneous Thorium-Argon. Almost 400 measurements with exposure time 
between 10 and 15 sec were obtained over a duration of 6.8 hours. The SNR ranged from
300 to 500, {\modiff and a computed} RV photon noise was in the range 55 - 90  {\cms}. Figure~\ref{figastero} shows the radial velocities of Procyon. The dispersion is equal to 3.14 {\ms}, but one may see clearly in the RV series oscillation modes with periods close to 17 mn.  
Figure~\ref{figastero} shows the amplitude spectrum of the radial velocities. The p-modes are visible as expected 
around 1 mHz. A mean noise level of 10.7 {\cms}  was computed in the amplitude spectrum in the frequency range 
3 - 9 mHz. Considering our 396 RV measurements, the corresponding velocity noise in the 
time series is 120 {\cms}. A previous run made with SOPHIE on Procyon by Mosser et al. (\cite{mosser08}) led to 220 {\cms}. A comparable sequence made on HARPS by Bouchy et al. (\cite{bouchy04}) led to a noise of 127 {\cms}. We also note that there is no significant signal at high frequency due to periodic error in the guiding system as it was observed on HARPS at 6 mHz (Bazot et al. \cite{bazot07}) before the installation of a tip-tilt system in 2011.

\subsubsection{Asteroseismology of Eta Cas}

Two sequences of RV measurement were made on the solar twin Eta Cas (HD\,4614)
using both mode HR+ and HE+ with simultaneous Thorium-Argon. With the HR mode, 250 measurements with exposure time between 30 and 45 sec were obtained over a duration of six hours. The SNR ranged from 130 to 220, {\modiff and a computed} RV photon noise was in between 80 and 140  {\cms}. 
With the HE+ mode, 270 measurements with an exposure time of 30 sec were obtained over a duration of 5.2 hours. 
The SNR ranged from 200 to 320, which led to an estimated RV photon noise between 60 and 100  {\cms}. Figure~\ref{figastero} shows the radial velocities of Eta Cas. The dispersion is equal to 1.8 and 1.9 {\ms} for the HE and HR modes, respectively. A slight drift of a few {\ms} appears in both time series, which is not understood. Figure~\ref{figastero} shows the amplitude spectrum of the radial velocities (HE and HR together). The p-modes are visible as expected around 3 mHz (corresponding to $\sim$5 minutes). A mean noise level of 8.7 {\cms}  was computed in the amplitude spectrum in the frequency range 4.5 - 6 mHz. Considering our 520 RV measurements, the corresponding velocity dispersion in the time series is 112 {\cms}. A tentative detection of p-modes close to 3 mHz on this target was claimed by Martic et al. (\cite{martic01}) based on six nights of ELODIE observations.

\begin{table}
\caption{Doppler helio- and asteroseimology with SOPHIE+}             
\label{tableastero}      
\centering                         
\begin{tabular}{cccccc}        
\hline\hline                 
Name & Spectral & mv & $\sigma_{RV}$  & $\sigma_{TF}$ & $\sigma_{noise}$  \\    
 & type  &   & [{\cms}]  &  [{\cms}]  &   [{\cms}]  \\    
\hline                        
SUN   & G2V   &  -   &  100   & 5.4 &  43  \\
HD139324 & G5IV-V & 7.48  & 235  & 44.0  & 182  \\
Procyon & F5IV-V    &  0.35    & 314  & 10.7  &  120  \\
Eta Cas   & G3V    & 3.45   & 185  & 8.7 &  112 \\
\hline                                   
\end{tabular}
\end{table}

\subsection{Follow-up observations of known low-mass exoplanets}

\begin{table*}
  \centering 
  \caption{Known low-mass exoplanets measured with SOPHIE+.}
  \label{table_known_planets}
\begin{tabular}{l|cc|ccccc|ccc}
\hline\hline
			& \multicolumn{2}{|c|}{Host star}   &	 \multicolumn{5}{|c|}{Published orbit}  & \multicolumn{3}{c}{SOPHIE+} \\
Planet		& Spectral   & $m_V$  & $m_p\sin i$	   & Period & $K$  &  $\sigma_\mathrm{O-C}$ & Reference   & $\sigma_\mathrm{O-C}$ & time span & N$_{obs}$ \\
			& type	&	& [M$_{\oplus}$] & [days]	 & [{\ms}] &    [{\ms}] &               & [{\ms}]                             &  [days]    \\
\hline
Gl\,436b		& M3.5V   & 10.6 & 	23  &  2.644 & 18.3  & 3.9 &  Maness et al.~(\cite{maness07}) & 2.1 & 210 & 6   \\
HD\,190360c	& G7IV-V  & 5.7   & 	19  & 17.111 & 4.8  &  2.5  &  Wright et al.~(\cite{wright09})	    &  1.0 & 22  & 15 \\
HD\,219828b	& G0IV      &  8.0  &  	20  &  3.833  & 7.0  & 1.7  &	Melo et al.~(\cite{melo07})             &  2.0 & 30 & 13  \\
HD\,7924b	& K0V       &   7.2  &  	9    &  5.398 & 3.9  &	2.8  & Howard et al.~(\cite{howard09})   &  1.7  & 11 & 9	\\
\hline
\end{tabular}
\end{table*}

\begin{figure*}
\begin{center}
\includegraphics[scale=0.45]{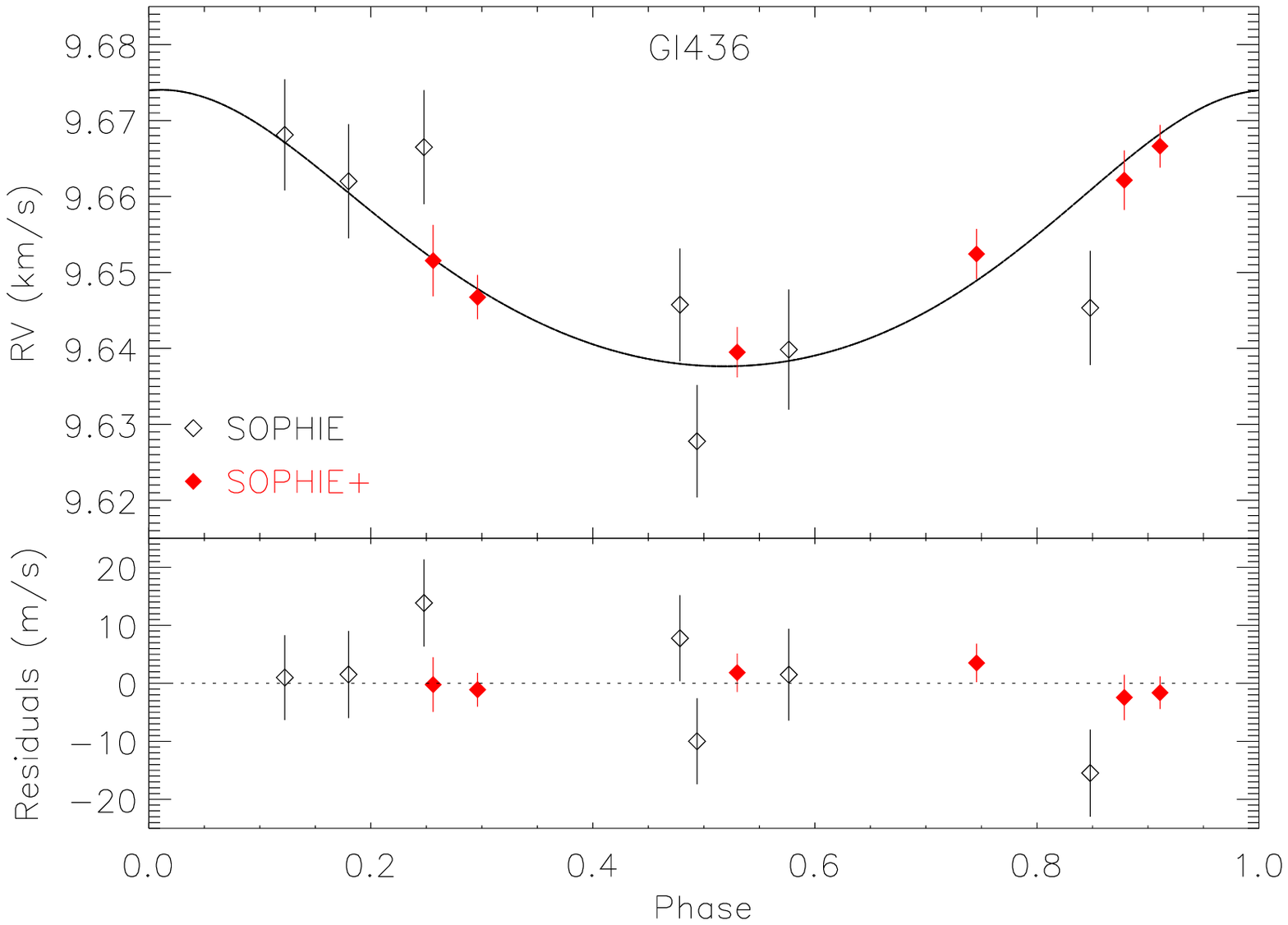}
\includegraphics[scale=0.45]{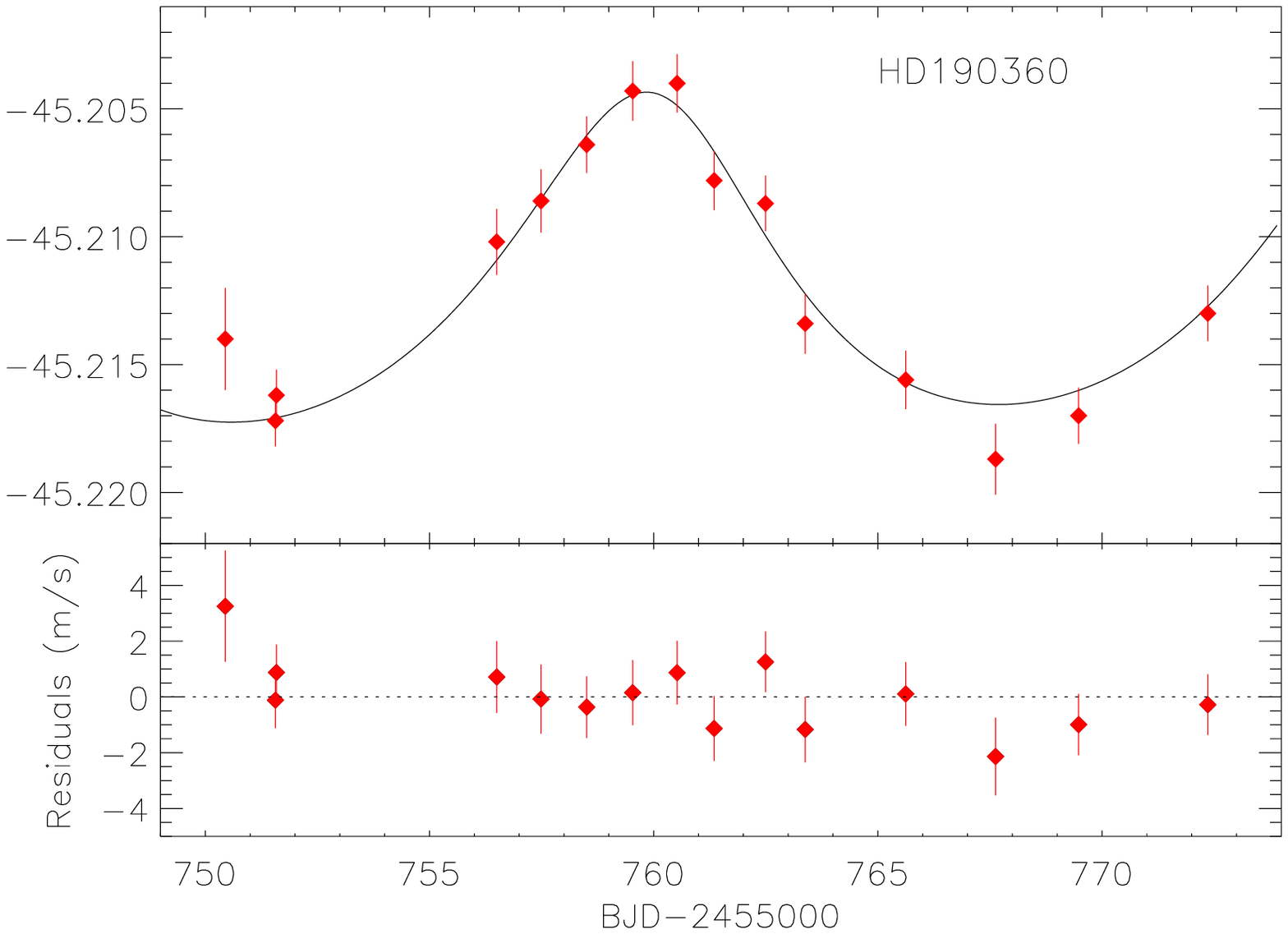}
\includegraphics[scale=0.45]{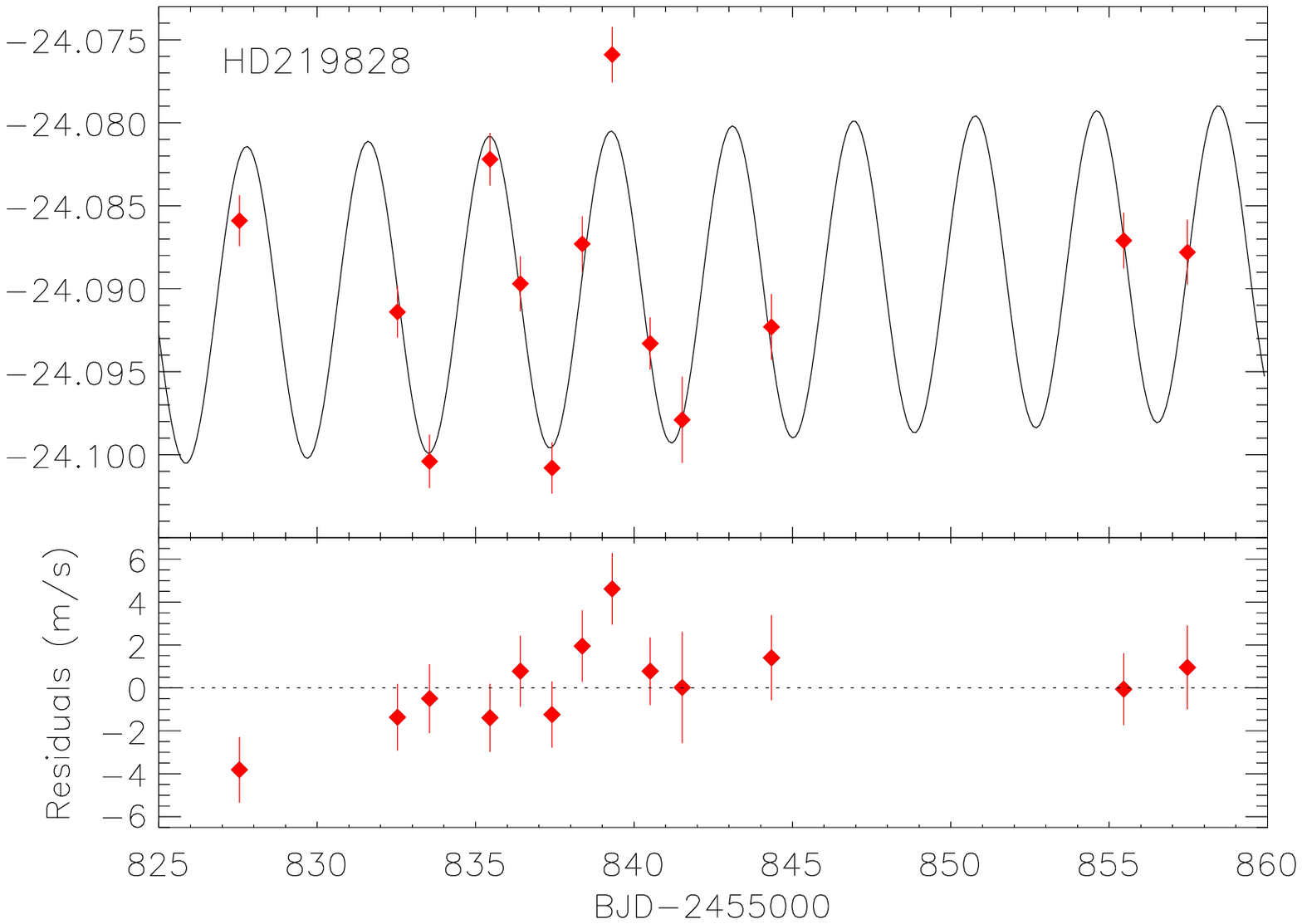}
\includegraphics[scale=0.45]{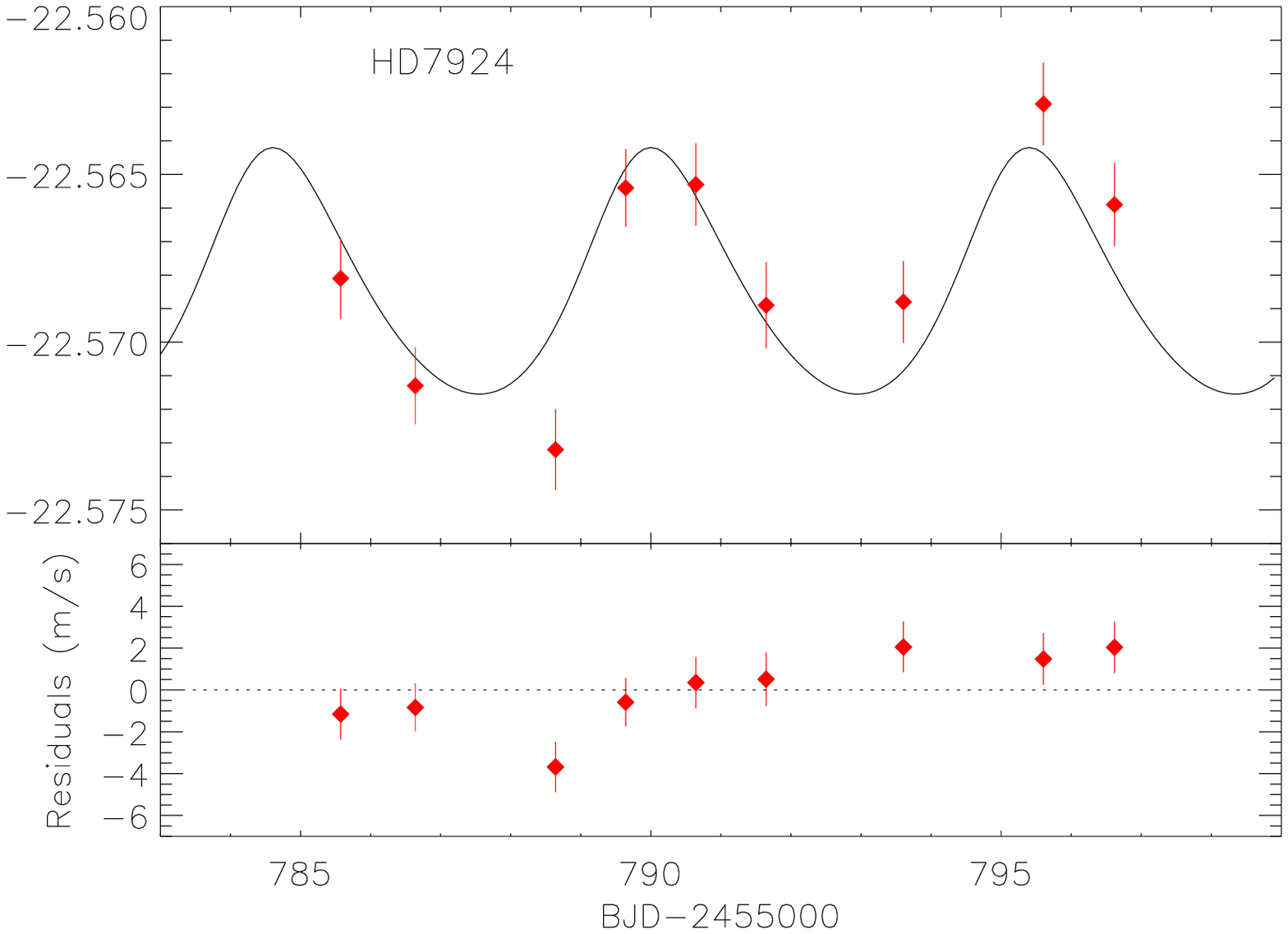}
\caption{Known low-mass exoplanets host stars measured with SOPHIE+.
\textit{Upper left panel:} phase-folded radial velocity curve of Gl\,436. The model includes 
the 2.6-day Keplerian.
Residuals of SOPHIE+ data (red, filled diamonds) are less dispersed than 
those obtained without the octagonal fiber (black, open diamonds).
\textit{Upper right panel:} SOPHIE+ radial velocities of HD\,190360 as a function of time. 
The model includes the 17.1-day Keplerian and the long-period planet, which produces 
a $\sim1$-{\ms} drift on the 3-week span displayed here.
\textit{Lower left panel:} SOPHIE+ radial velocities of HD\,219828 as a function of time. 
The model includes the 3.8-day Keplerian, as well as the long-period planet which produces 
a $\sim2$-{\ms} drift on the 1-month span displayed here.
\textit{Lower right panel:} SOPHIE+ radial velocities of HD\,7924 as a function of time. 
The model only includes the 5.4-day Keplerian.
}
\label{fig_known_planets}
\end{center}
\end{figure*}

To validate the capability of SOPHIE+ to detect low-mass exoplanets, we measured four stars with 
known super-Earth or Neptune-mass planets on short periods: Gl\,436, HD\,190360, HD\,219828, 
and HD\,7924. These four targets were observed in HR+ mode with simultaneous Thorium-Argon. 
The parameters of these systems are reported in Table~\ref{table_known_planets} and their SOPHIE+ orbits are plotted in Fig.~\ref{fig_known_planets}, {\modif together with the Keplerian fits described below, obtained 
using a Levenberg-Marquardt algorithm}.

\subsubsection{Gl436b}	

The hot Neptune orbiting the M-dwarf Gl\,436 was discovered by Butler et al.~(\cite{butler04}); Gillon et al.~(\cite{gillon07}) showed it was transiting its star each 2.6 days. Seven 1800 sec observations of that target were 
secured with SOPHIE between 2008 February and 2011 April, without simultaneous Thorium-Argon and before the 
implementation of octagonal fibers. Three 1800 sec exposures in a few nights were obtained with SOPHIE+ in June 2011, then three others in February 2012. Other radial velocities available in the literature include 
the 59 HIRES/Keck measurements obtained over 6.5 years (Maness et al.~\cite{maness07}).

Following Maness et al.~(\cite{maness07}), we fitted the SOPHIE and HIRES data with a Keplerian model. We fixed the eccentricity to the value $e=0.14$ found by Demory et 
al.~(\cite{demory07}) from transit and occultation timing. The parameters we derived for the orbit agree with those already published. The standard deviation of the residuals of the HIRES data is 
$\sigma_\mathrm{O-C}=3.9$\,{\ms}, similar to the 4.3 {\ms} dispersion reported by Maness et al.~(\cite{maness07}). This is slightly larger than the expected error bars on individual radial velocities. The additional dispersion was attributed to a stellar jitter of the order of 2~{\ms} by Maness et al.~(\cite{maness07})and Demory et al.~(\cite{demory07}).

The residuals of the SOPHIE data without and with octagonal fibers show 
dispersions $\sigma_\mathrm{O-C}=9.2$~{\ms} and 2.1~{\ms}, respectively. 
This shows the real improvement due to the octagonal fibers (see upper  
left-hand panel of Fig.~\ref{fig_known_planets}). The 2.1 {\ms} dispersion is slightly 
smaller than the 3.5 {\ms} typical photon noise uncertainty on individual measurements. 
This could be due to the small number (six) of measurements during that 210-day 
time span. However,  the SOPHIE+ data do not suggest significant stellar 
jitter on that target. The HIRES data show a larger dispersion, but they cover a 
longer time span. Considering only 30-day sequences, the HIRES data typically 
show $\sigma_\mathrm{O-C}=3.6$~{\ms}, not significantly smaller than the full 
6.5-year sequence. 

\subsubsection{HD\,190360c}

The G7IV-V, inactive star HD\,190360 harbors a giant planet on an 8-year orbit and a 
19 M$_{\oplus}$ planet on a 17.1-day period. That low-mass planet, HD\,190360c, 
was detected by Vogt et al.~(\cite{vogt05}) with HIRES. Wright et al.~(\cite{wright09}) 
provide the latest version of the HIRES measurements, including a total of 107 
radial velocities on a 12-year time~span.

We obtained 15 SOPHIE+ measurements of HD\,190360 over 22 days in July 2011 
in order to sample the entire period of the low-mass planet. The exposure times 
were typically ten minutes. The upper right-hand panel of Fig.~\ref{fig_known_planets} displays those
new measurements and a two-Keplerian fit {\modif of the SOPHIE+ data alone}, the long period being fixed to the solution of Wright et al.~(\cite{wright09}). The dispersion of the residuals is 1.0\,{\ms}, in good agreement 
with the typical error bars of the individual measurements. The solution 
for HD\,190360c agrees with that of the literature, with the exception of the semi-amplitude:
we found $K=6.3\pm0.3$\,{\ms}, which is 2.5\,$\sigma$ larger than the value obtained by 
Wright et al.~(\cite{wright09}). The {\modif corresponding} projected mass is $m_p\sin i = 24.8 \pm 2.4 $\,M$_{\oplus}$.
{\modif Additional SOPHIE+ observations would be needed before suggesting a significant revision
of the mass.} By forcing the semi-amplitude to the value of Wright et al.~(\cite{wright09}) {\modif 
or nearly identically by fitting simultaneously the SOPHIE+ and HIRES data}, 
the SOPHIE+ residuals show a slightly increased dispersion from 1.0\,{\ms} to 1.4\,{\ms}. 
By comparison, the HIRES dataset shows a larger dispersion around the fit, 
$\sigma_\mathrm{O-C}=2.5$\,{\ms}. The dispersion remains similar when shorter sequences of the HIRES data are used in order to match the shorter time span of the SOPHIE+ observations. Thus the dispersion seen in the HIRES
data does not seem to be due to low-frequency variations.

\subsubsection{HD\,219828b}

Melo et al.~(\cite{melo07}) published the detection of a 20 M$_{\oplus}$ planet on a 3.8-day 
period around the G0IV, inactive star HD\,219828 from HARPS measurements. 
Long-term RV variations were also identified in addition to the variations due to the hot Neptune HD\,219828b. We secured 13 SOPHIE+ measurements during a time span of one month in fall 2011, with 15- to 
20-minute exposures. According to the ongoing HARPS follow-up of that target, the long-term trend 
in that month was below 2\,{\ms} (S.~Udry, private communication).

The lower left-hand panel of Fig.~\ref{fig_known_planets} shows the Keplerian fit of the hot Neptune, 
together with the {\modif fixed} trend. {\modif Fitted alone,} the SOPHIE+ data show a 2.0 {\ms} dispersion around that fit, in good 
agreement with the error bars on individual measurements. The derived parameters agree 
with those derived by Melo et al.~(\cite{melo07}), except for the semi-amplitude. We found 
$K=9.5\pm0.5$\,{\ms}, i.e. 3.5\,$\sigma$ larger than the value from Melo et al.~(\cite{melo07}).
This would imply a projected mass around $m_p\sin i = 27$\,M$_{\oplus}$. 
Melo et al.~(\cite{melo07}) obtained a 1.7 {\ms} dispersion of the residuals of the HARPS 
data, which span 1.3~years. The HARPS data seem to be slightly more accurate than the 
SOPHIE+ data. By forcing the SOPHIE+ fit to match the solution from Melo et al.~(\cite{melo07}),
we increased the residuals dispersion from 2.0 to 2.6\,{\ms}

\subsubsection{HD\,7924b}

HD\,7924 is a K0V star that is slightly active ($\log{R'_\mathrm{HK}} = -4.89$). 
A seven-year follow up of that target with HIRES has revealed the presence of a super 
Earth in orbit around it (Howard et al.~\cite{howard09}). HD\,7924b has a 9.26 M$_{\oplus}$
projected mass and a 5.4-day orbit. We present here nine SOPHIE+ measurements 
performed on that target over 11~days in late summer 2011. Exposure times were around 
15~minutes. The lower right-hand panel of Fig.~\ref{fig_known_planets} shows that SOPHIE+ 
can detect such a super-Earth. {\modif The Keplerian fit includes the SOPHIE+ and HIRES data and agrees with the orbit parameters from Howard et al.~(\cite{howard09}). The SOPHIE+ residuals show a dispersion 
of 1.7\,{\ms}}. This is smaller than the 
dispersion of the whole HIRES dataset, which presents a dispersion of 2.8\,{\ms} over 
seven years. However, when considering sections of the HIRES dataset with short time 
spans similar to that of SOPHIE+, we found no significant differences between SOPHIE+
and HIRES accuracies for this target. {\modif A linear trend could be present in the residuals of the SOPHIE+ 
data, and additional measurements are required to confirm it.}

\section{Discussions and conclusions}

The implementation of an octagonal fiber in the fiber link of the SOPHIE spectrograph 
improves by a factor $\sim$6 its RV precision. The main limitation of SOPHIE, related 
to its sensitivity to pupil illumination changes due to imperfect fiber scrambling, is now strongly 
reduced, thanks to octagonal fiber. The typical precision of SOPHIE+ is now in the range 1-2 {\ms} on standard solar-type stars observed with the HR+ mode. The systematic effects due to guiding decentering, telescope defocusing, 
atmospheric dispersion, and seeing variations have now vanished under the 1-2 {\ms} level.   

Our high-cadence RV sequences illustrate the gain of SOPHIE+ for Doppler asteroseismology 
compared to the previous fiber link configuration. The performances of SOPHIE+ 
for asteroseismology are now close to the photon noise uncertainty and close to those of HARPS.

\begin{figure}
   \centering
   \includegraphics[width=8cm]{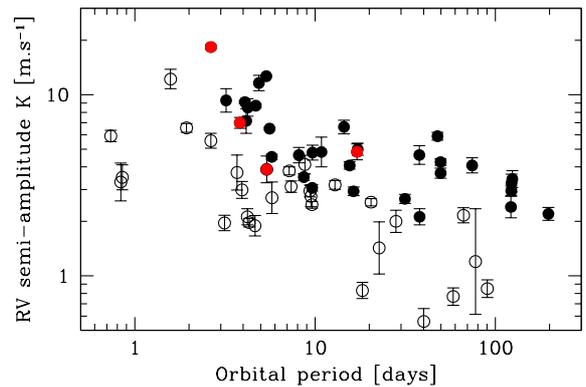}
      \caption{Radial velocity semi-amplitude $K$ as function of orbital period for known low-mass exoplanets ($m$\,sin\,$i$ $\le$ 0.1 {\Mjup}). Black dots and white dots  correspond to exoplanets with {\modif minimum} masses above and below 10 M$_{Earth}$ respectively. The four red dots correspond 
      to the exoplanets observed with SOPHIE+.}
         \label{figplanet}
   \end{figure}

Our scientific validation demonstrates the capability of SOPHIE+ to detect short-period, 
low-mass planets with RV semi-amplitude of a few {\ms}. On time scales of 
a few tens of days studied here, SOPHIE+, which is mounted at a 2 m class telescope, is shown to have an accuracy of 2\,{\ms} or even better on bright targets. Its {\modif RV precision} is similar to that of the HIRES spectrograph on the 10.2 m Keck telescope and approaches that of HARPS at the 3.6 m ESO telescope in La Silla. 
Figure~\ref{figplanet} shows the RV semi-amplitude $K$ as function of the orbital period for  
known low-mass exoplanets with masses smaller the 0.1 {\Mjup}\footnote{from http://exoplanets.org}. 
Our results show that SOPHIE+ can now efficiently detect exoplanets with semi-amplitudes down to 3 {\ms} and with periods up to 20 days. This corresponds to 50\% of known low-mass exoplanets. 

For a non-rotating K-type star of $m_v=8$, a SNR per pixel at 550 nm of 150 is obtained in 20 mn with the HR mode.  
With such a SNR, the photon noise uncertainty is 1 {\ms}. For exoplanet searches, an exposure time of 15-20 mn is 
required to average the p-mode oscillations (e.g., Bouchy et al. \cite{bouchy05}). Even with a 2 m class telescope, 
a large search program for low-mass exoplanets at the level of 1-2 {\ms} precision can be conducted 
on a significantly large number of stars.  

For the follow-up of small-size transiting candidates of the space mission {\it CoRoT} and {\it Kepler}, 
SOPHIE+ is limited to the brightest targets with $m_v \le 11$. For future space missions 
like TESS or PLATO, which are devoted to searching for terrestrial transiting planets around bright stars, 
instruments similar to SOPHIE+ will be able to play a key role, as they already do for giant transiting planets.  

This first on-sky demonstration of the gain provided by octagonal fibers for 
high-precision fiber-fed spectrographs will benefit next-generation instruments 
like HARPS-North {\modif (Cosentino et al. \cite{cosentino12})}, ESPRESSO {\modif (Pepe et al. \cite{pepe10})} 
and SPIRou {\modiff (Artigau et al. \cite{artigau11})}. 
{\modif The use of octagonal fibers does not reduce or relax the importance of the opto-mechanical 
stability of the spectrograph itself. For spectrographs using the simultaneous wavelength calibration 
method with two fibers, the spectrograph should be sufficiently stabilized to have no significant relative drift 
between the two fibers.} Furthermore, for spectrographs using 
the iodine cell method, a high stabilization of the PSF may be critical and an octagonal 
fiber link may certainty help improve the iodine cell technique. 

Future improvements to SOPHIE+ could consist of implementing a piece of octagonal fiber 
in the HR+ mode after the double scrambler in order to perfectly scramble both near- and far field of the output 
beam of the fiber. A new calibration unit is presently under development to better control the calibration lamp 
and monitor the long-term evolution of Thorium-Argon lamp. This new calibration unit 
will permit a Fabry-Perot Etalon or a laser frequency comb to be added in order to derive the instrumental drift more precisely than with the Thorium-Argon lamp.

\begin{acknowledgements}
We thank the technical team at the Observatoire de Haute-Provence for their support with the \textit{SOPHIE} instrument and the 1.93-m telescope and, in particular, for the essential work of the night assistants. Financial support from the ÒProgramme national de plan\'etologieÓ (PNP) of the CNRS/INSU, France, is gratefully acknowledged. We also acknowledge support from the French National Research Agency (ANR-08-JCJC-0102-01). 
\end{acknowledgements}

\end{document}